\documentclass{ws-ijmpa}
\usepackage{amssymb,amsmath,latexsym}
\usepackage{graphicx}
\usepackage{hyperref}

\newcommand\rf[1]{(\ref{eq:#1})}
\newcommand\lab[1]{\label{eq:#1}}
\newcommand\nonu{\nonumber}
\newcommand\br{\begin{eqnarray}}
\newcommand\er{\end{eqnarray}}
\newcommand\be{\begin{equation}}
\newcommand\ee{\end{equation}}

\newcommand\lb{\lbrack}
\newcommand\rb{\rbrack}

\newcommand\rrangle{\right\rangle}

\newcommand\llb{\left\lbrack}
\newcommand\rrb{\right\rbrack}

\renewcommand\({\left(}
\renewcommand\){\right)}
\newcommand\bv{\bigm\vert}               

\newcommand\bc{\begin{center}}
\newcommand\ec{\end{center}}

\newcommand\Tr{\mathop{\mathrm Tr}}                  
\newcommand\partder[2]{\frac{{\partial {#1}}}{{\partial {#2}}}}

\renewcommand\a{\alpha}
\renewcommand\b{\beta}

\renewcommand\d{\delta}

\newcommand\eps{\epsilon}
\newcommand\vareps{\varepsilon}
\newcommand\g{\gamma}
\newcommand\G{\Gamma}

\newcommand\h{\frac{1}{2}}
\renewcommand\k{\kappa}
\renewcommand\l{\lambda}
\renewcommand\L{\Lambda}
\newcommand\m{\mu}
\newcommand\n{\nu}

\renewcommand\O{\Omega}

\newcommand\vp{\varphi}
\renewcommand\P{\Phi}
\newcommand\pa{\partial}

\newcommand\pr{\prime}

\renewcommand\th{\theta}

\newcommand\wti{\widetilde}

\newcommand\cA{{\mathcal A}}

\newcommand\cH{{\mathcal H}}

\newcommand{\ct}[1]{\cite{#1}}
\newcommand{\bib}[1]{\bibitem{#1}}
\newcommand\PRL[3]{\textsl{Phys. Rev. Lett.} \textbf{#1}, #3 (#2)}
\newcommand\NPB[3]{\textsl{Nucl. Phys.} \textbf{B#1}, #3 (#2)}

\newcommand\PRD[3]{\textsl{Phys. Rev.} \textbf{D#1}, #3 (#2)}

\newcommand\PLB[3]{\textsl{Phys. Lett.} \textbf{#1B}, #3 (#2)}

\newcommand\AoP[3]{\textsl{Ann. of Phys.} \textbf{#1}, #3 (#2)}

\newcommand\MPLA[3]{\textsl{Mod. Phys. Lett.} \textbf{A#1}, #3 (#2)}

\newcommand\udot{\stackrel{.}{u}}

\newcommand\Adot{\stackrel{.}{A}}
\newcommand\Bdot{\stackrel{.}{B}}
\newcommand\Hdot{\stackrel{.}{H}}

\begin{document}

\markboth{E. Guendelman, E. Nissimov and S. Pacheva}
{Vacuum Structure and Gravitational Bags}

%
\catchline{}{}{}{}{}
%

\title{Vacuum Structure and Gravitational Bags Produced by Metric-Independent 
Spacetime Volume-Form Dynamics}

\author{Eduardo Guendelman}

\address{Department of Physics, Ben-Gurion University of the Negev\\
P.O.Box 653, IL-84105 ~Beer-Sheva, Israel \\
guendel@bgu.ac.il}

\author{Emil Nissimov and Svetlana Pacheva}

\address{Institute for Nuclear Research and Nuclear Energy, Bulgarian Academy
of Sciences\\
Boul. Tsarigradsko Chausee 72, BG-1784 ~Sofia, Bulgaria \\
nissimov@inrne.bas.bg, svetlana@inrne.bas.bg}

\maketitle

\begin{history}
\received{}
\revised{}
\end{history}

\vspace{-.1in}
\begin{abstract}
We propose a new class of gravity-matter theories, describing $R+R^2$ gravity
interacting with a nonstandard nonlinear gauge field system and a scalar
``dilaton'', formulated in terms of two different non-Riemannian volume-forms
(generally covariant integration measure densities) on the underlying spacetime
manifold, which are independent of the Riemannian metric. The nonlinear gauge field 
system contains a square-root $\sqrt{-F^2}$ of the standard Maxwell Lagrangian which
is known to describe charge confinement in flat spacetime. 
The initial new gravity-matter model 
is invariant under global Weyl-scale symmetry which undergoes a spontaneous
breakdown upon integration of the non-Riemannian volume-form degrees of freedom.
In the physical Einstein frame we obtain an effective matter-gauge-field Lagrangian of
``k-essence'' type with quadratic dependence on the scalar ``dilaton'' field kinetic
term $X$, with a remarkable effective scalar potential possessing two infinitely 
large flat regions as well as with nontrivial effective gauge coupling constants
running with the ``dilaton'' $\vp$. Corresponding to the each of the two flat regions
we find 
``vacuum'' configurations of the following types:
(i) $\vp = {\rm const}$ and a non-zero gauge field vacuum $\sqrt{-F^2}\neq 0$, 
which corresponds to a charge confining phase;
(ii) $X = {\rm const}$ (``kinetic vacuum'') and ordinary gauge
field vacuum $\sqrt{-F^2}=0$ which supports confinement-free charge dynamics.
In one of the flat regions of the effective scalar potential we also find:
(iii) $X = {\rm const}$ (``kinetic vacuum'') 
and a non-zero gauge field vacuum $\sqrt{-F^2}\neq 0$, which again corresponds to
a charge confining phase.
In all three cases the spacetime metric is de Sitter or Schwarzschild-de Sitter.
Both ``kinetic vacuums'' (ii) and (iii) can exist only within a
finite-volume space region below a de Sitter horizon. Extension to the whole
space requires matching the latter with the exterior region with a nonstandard
Reissner-Nordstr{\"o}m-de Sitter geometry carrying an additional constant
radial background electric field. As a result we obtain two classes of gravitational
bag-like configurations with properties, which on one hand partially parallel some of
the properties of the solitonic ``constituent quark'' model and, on the
other hand, partially mimic some of the properties of MIT bags in QCD phenomenology.
\end{abstract}


\ccode{PACS numbers: 04.50.Kd, 
11.30.Qc, 
98.80.Bp, 
95.36.+x 
}

\section{Introduction}

To understand the basic features of hadronic physics we are unavoidably lead 
to the existence of different phases of the gauge theory.
Furthermore, these different phases are apparently associated with different values
of the vacuum energy density, as described for example in the {\em MIT bag model} 
of hadrons \ct{MIT-bag-1,MIT-bag-2}. Inside the MIT bag quarks and gluons propagate 
almost freely and there the vacuum energy density is big defining the so called bag 
constant.Outside the bag there is no propagation of either quarks or gluons, and in the 
MIT bag model the outside vacuum energy density is set to zero 
(any choice for the outside energy density is possible, if we ignore gravity, 
since while ignoring gravity only the difference of the energy densities inside and 
outside the bag is of significance). For a vacuum state $p=-\rho$, where 
the vacuum pressure inside the bag is negative while zero outside, an empty bag 
therefore tends to implode. When the bag is filled with particles, the positive
pressure of the particles stabilizes the bubble at a certain radius.

It is interesting that a similar ``two phase'' structure defined via two
vastly different scales of the vacuum energy density does also appear in cosmology.
Indeed, according to the most accepted scenario of the early universe -- the 
inflation picture \ct{early-univ-1}${}^{-}$\ct{early-univ-6}, there was at the beginning 
a large vacuum energy density. Together with this, for a description of the present 
slowly accelerated phase of the universe \ct{slow-accel-1}${}^{-}$\ct{slow-accel-5} 
one employs a small vacuum energy density.

A scenario of continuously connecting an inflationary phase to a slowly
accelerating universe through the evolution of a single scalar field -- the
so called {\em ``quintessential inflation''} scenario --  has been first 
proposed in Ref.~\refcite{peebles-vilenkin}, which triggered active further development
(models based on generalized $F(R)$ gravity \ct{starobinsky-2}; based on 
the k-essence \ct{k-essence-1}${}^{-}$\ct{k-essence-4} framework -- 
see Ref.~\refcite{saitou-nojiri}; based on the ``variable gravity''"mode\ct{wetterich}
and containing an extensive list of references to earlier work on the topic -- 
see Ref.~\refcite{murzakulov-etal-1}, see also Ref.~\refcite{murzakulov-etal-2}).

In the cosmological context we have been able to construct models providing 
a unified scenario where both an inflation and a slowly accelerated phase for the
universe can appear naturally from the existence of two infinitely large flat regions
in the effective scalar field potential with vastly different scales which we derive
systematically from a Lagrangian action principle \ct{emergent,quintess}. 
Namely, we have constructed a new kind of globally Weyl-scale invariant gravity-matter 
action within the first-order (Palatini) approach formulated in terms of two 
different non-Riemannian volume-forms (generally covariant integration measure
densities on the pertinent spacetime manifold independent of the Riemannian metric). 
The principal feature is the
requirement of global Weyl-scale invariance and the choice of different scaling 
properties of the two non-Riemannian measures (volume elements) which dictates the
precise form of the terms in the action. In this new theory there is a single scalar
field with kinetic terms coupled to both non-Riemannian 
measures, whereas the standard Einstein-Hilbert term $R$ and the $R^2$-term couple
each to a different non-Riemannian measure.

Global Weyl-scale invariance is spontaneously broken upon solving part 
of the equations of motion corresponding to the auxiliary antisymmetric 
tensor gauge fields of maximal rank
defining the two non-Riemannian measures -- due to the appearance of two 
arbitrary dimensionful integration constants. The latter produce a remarkable
effect on the resulting physical Einstein-frame theory \ct{emergent,quintess} 
-- we find there an effective k-essence \ct{k-essence-1}${}^{-}$\ct{k-essence-4} 
type of theory, where the effective scalar field potential has 
{\em two infinitely large flat regions} corresponding to the two accelerating phases 
of the universe -- one for large negative values of the scalar field with a
very large height corresponding to the early universe, and another one for
large positive values of the scalar field with a very low height corresponding to the 
universe of the present epoch.

Since the construction was based on geometrical and symmetry considerations, 
which are very general, one may think that a similar model can be constructed 
for a different physical application, \textsl{i.e}, the phases of a gauge theory.
As we will see here, it is indeed possible to obtain a phase structure of
confinement and deconfinement related to what the MIT bag model suggests.

To this end we will couple the above new type of gravity-matter theory
defined in terms of the two different non-Riemannian measures and containing
$R^2$ gravity term to a special kind of nonstandard nonlinear gauge field
model (for the analogous situation in the less general case of
gravity-matter models with one non-Riemannian and one standard Riemannian
integration measures, see Ref.~\refcite{GG-0}).

Namely, let us consider Abelian gauge fields whose Lagrangian contains both the
standard Maxwell Lagrangian ($\sim F^2 \equiv F_{\m\n}F^{\m\n}$) as well as the 
nonstandard square-root of the latter ($\sim \sqrt{-F^2}$) coupled to the two 
different non-Riemannian measures in a globally Weyl-scale invariant form. In flat
space-time the $\sqrt{-F^2}$-term is known to describe dynamics of charge 
confinement \ct{GG-1}${}^{-}$\ct{GG-6} related to the non-zero vacuum value of 
$\sqrt{-F^2}$. The latter is an explicit realization of an earlier proposal by 
`t Hooft \ct{thooft-1,thooft-2} who argued that the energy density of electrostatic
field configurations in the low-energy description of confining quantum gauge 
theories must be a linear function of the electric displacement field in the
infrared region (the latter appearing as a quantum ``infrared counterterm''). 
Below, in \textsl{Appendix B} we extend the flat spacetime proof in Ref.~\refcite{GG-2} 
about the charge confining property of the $\sqrt{-F^2}$-term to the case of
curved static spherically-symmetric spacetimes.

For further interesting properties of gravity-matter theories involving the
``square-root'' Maxwell term $\sqrt{-F^2}$ (black holes with confining electric 
potential, new mechanism for dynamical generation of cosmological constant, 
charge-hiding and charge-confining via ``tube-like'' wormholes), 
see Refs.~\refcite{our-sqroot-1}-\refcite{our-sqroot-3}.

Let us remark that one could start with the non-Abelian version of the 
nonlinear gauge field system with $\sqrt{- \Tr\bigl(F_{\m\n}F^{\m\n}\bigr)}$.
Since we will be interested in static spherically symmetric solutions,
the non-Abelian theory effectively reduces to an Abelian one as pointed out
in Refs.~\refcite{GG-1}-\refcite{GG-6}.

In the present context after including the coupling of the
non-Riemannian-measures-modified gravity-matter theory to the nonlinear gauge
field with $\sqrt{-F^2}$, in the pertinent physical Einstein frame we obtain an 
effective matter Lagrangian again of ``k-essence'' type with quadratic dependence 
on the $\vp$ kinetic term $X$ of the scalar ``dilaton'' field $\vp$, 
with a remarkable effective 
scalar potential possessing two infinite flat regions with different energy
scales. In addition we get a nontrivial coupling of the nonlinear gauge
field to the ``dilaton'' kinetic term $X \sqrt{-F^2}$. 
All terms are multiplied by nontrivial ``dilaton''-dependent 
coefficient functions, including nontrivial effective gauge coupling constants running 
with $\vp$. An important observation is their ``flatness'' (constancy w.r.t. 
running $\vp$) in both infinitely large flat regions of the effective potential. 

We study the static spherically symmetric ``vacuum'' configurations corresponding to
each of the two flat regions. In all cases the gravitational part is de
Sitter type (de Sitter or Schwarzschild-de Sitter) with effective cosmological constant 
whose value is determined by the height of the total effective ``dilaton'' potential in each of 
the flat regions. The latter includes, apart from the purely scalar ``dilatonic'' ones, 
also additional contributions due to (possible) non-zero vacuum values of $\sqrt{-F^2}$ and $X$.

The static spherically symmetric ``vacuum'' configurations of 
the ``dilaton'' $\vp$ and the nonlinear gauge field are of the following types: 

(i) $\vp = {\rm const}$, \textsl{i.e.}, $X=0$ and a non-zero gauge field vacuum 
$\sqrt{-F^2}\neq 0$, the latter corresponding to a confining phase -- these
solutions exist in both flat regions of the effective scalar potential
(the one for large negative values of $\vp$ 
and the other one for large positive values of $\vp$)

(ii) $X = {\rm const}$ (``kinetic vacuum'') and ordinary gauge
field vacuum $\sqrt{-F^2}=0$, which supports {\em confinement-free} charge dynamics
-- this solution exists both in the flat region of the effective scalar potential
for large positive values of $\vp$ as well as for a special value of one of the
scalar potential's parameters also in the flat region for large negative values of $\vp$.

(iii) $X = {\rm const}$ (``kinetic vacuum'') and a non-zero gauge field vacuum 
$\sqrt{-F^2}\neq 0$, which again corresponds to a confining phase -- this
solution exist in the flat region of the effective scalar potential
for large negative values of $\vp$ for generic scalar potential's parameter values.

An important point here is that both ``kinetic vacuums'' (ii) and (iii) 
do not represent themselves genuine vacuum configurations, since they are defined 
only within a finite-volume space region below the de Sitter horizon. In
order to obtain a well-defined static spherically symmetric configuration over 
the whole spacetime we need to match at the de Sitter horizon  the ``kinetic vacuums''
(ii) and (iii) living in the interior de Sitter region to the exterior region 
with a nonstandard Reissner-Nordstr{\"o}m-de Sitter geometry which carries an
additional constant radial background electric field. Studying the vacuum energy
densities inside and outside the de Sitter horizon shows that the inside
energy density is higher than the outside one. Thus, the fully
extended to the whole spacetime ``kinetic vacuums'' (ii) and (iii) represent
gravitational bag-like configurations where:

(a) The type (ii) 
gravitational bag mimics some
of the properties of the MIT bag model \ct{MIT-bag-1,MIT-bag-2} -- finite
volume space region with {\em deconfinement} and large energy density versus
infinite volume exterior region with {\em confinement} and low energy density;

(b) Both type (ii) and type (iii) 
gravitational bags resemble some of the properties of the solitonic ``constituent
quark'' model \ct{const-quark} -- they are charged 
and carry ``color'' flux to infinity.

The plan of the paper is as follows. In the next Section 2 we describe in
some detail the general formalism for the new class of gravity-matter
systems defined in terms of two independent non-Riemannian integration
measures. In Sections 3 and 4 we describe the properties of the two flat regions of the 
Einstein-frame effective scalar potential and derive the static spherically symmetric 
vacuum configurations. In Section 5 we construct static spherically symmetric
solutions representing gravitational bag-like configurations.
We conclude in Section 6 with some discussions. In \textsl{Appendix A} we briefly
outline the canonical Hamiltonian treatment of the modified gravity-matter
models with two non-Riemannian spacetime volume-forms, which elucidates the
physical meaning of the auxiliary fields defining the non-Riemannian volume-forms.
In \textsl{Appendix B} following Ref.~\refcite{GG-2}  we show that the
presence of the ``square-root'' Maxwell term $\sqrt{-F^2}$ generates confining
effective potential between quantized charged fermions in static spherically
symmetric spacetimes.

\section{Gravity-Matter System Coupled to Charge-Confining Nonlinear Gauge Field 
-- A Formalism With Two Independent Non-Riemannian Volume-Forms}
We shall consider the following nonstandard gravity/nonlinear-gauge-field/matter 
system with an action of the general form involving two independent non-Riemannian 
integration measure densities generalizing the models studied in 
Refs.~\refcite{quintess,emergent} (for simplicity we will use units where the Newton 
constant is taken as $G_{\rm Newton} = 1/16\pi$):
\be
S = \int d^4 x\,\P_1 (A) \Bigl\lb R + L^{(1)} \Bigr\rb +
\int d^4 x\,\P_2 (B) \Bigl\lb L^{(2)} + \eps R^2 
+ \frac{\P (H)}{\sqrt{-g}}\Bigr\rb \; .
\lab{TMMT+GG}
\ee
Here the following notations are used:

\begin{itemize}
\item
$\P_{1}(A)$ and $\P_2 (B)$ are two independent non-Riemannian volume-forms, 
\textsl{i.e.}, generally covariant integration measure densities on the underlying
space-time manifold:
\be
\P_1 (A) = \frac{1}{3!}\vareps^{\m\n\k\l} \pa_\m A_{\n\k\l} \quad ,\quad
\P_2 (B) = \frac{1}{3!}\vareps^{\m\n\k\l} \pa_\m B_{\n\k\l} \; ,
\lab{Phi-1-2}
\ee
defined in terms of field-strengths of two auxiliary 3-index antisymmetric
tensor gauge fields. $\P_{1,2}$ take over the role of the standard
Riemannian integration measure density 
$\sqrt{-g} \equiv \sqrt{-\det\Vert g_{\m\n}\Vert}$ in terms of the space-time
metric $g_{\m\n}$.
\item
$R = g^{\m\n} R_{\m\n}(\G)$ and $R_{\m\n}(\G)$ are the scalar curvature and the 
Ricci tensor in the first-order (Palatini) formalism, where the affine
connection $\G^\m_{\n\l}$ is \textsl{a priori} independent of the metric $g_{\m\n}$.
Note that in the second action term we have added a $R^2$ gravity term
(again in the Palatini form). Let us recall that $R+R^2$ gravity within the
second order formalism (which was also the first inflationary model) was originally
proposed in Ref.~\refcite{starobinsky}.
\item
$L^{(1,2)}$ denote two different Lagrangians of a single scalar matter field 
(``dilaton'') and of a Abelian gauge field potential $A_\m$ of the form:
\br
L^{(1)} = -\h g^{\m\n} \pa_\m \vp \pa_\n \vp - V(\vp) 
- \frac{f_0}{2}\sqrt{- F^2(g)} \quad ,\quad
V(\vp) = f_1 \exp \{-\a\vp\} \; ,
\lab{L-1} \\
L^{(2)} = -\frac{b}{2} e^{-\a\vp} g^{\m\n} \pa_\m \vp \pa_\n \vp + U(\vp) 
 - \frac{1}{4e^2} F^2(g) \quad ,\quad U(\vp) = f_2 \exp \{-2\a\vp\} \; ,
\lab{L-2}
\er
where: 
\be
F^2(g) = F_{\m\n} F_{\k\l} g^{\m\k} g^{\n\l} \quad ,\quad
F_{\m\n} = \pa_\m A_\n - \pa_\n A_\m \; .
\lab{F-def}
\ee
Here $\a, f_1, f_2$ are dimensionful positive parameter, whereas $b$ is a
dimensionless one. The choice of the scalar potentials in \rf{L-1}-\rf{L-2} is
similar to the choice in Ref.~\refcite{TMT-orig-1}.
\item
$\P (H)$ indicate the dual field strength of a third auxiliary 3-index antisymmetric
tensor gauge field:
\be
\P (H) = \frac{1}{3!}\vareps^{\m\n\k\l} \pa_\m H_{\n\k\l} \; ,
\lab{Phi-H}
\ee
whose presence is crucial for non-triviality of the model. 
\end{itemize}

Concerning the explicit form of the non-Riemannian integration measure
densities \rf{Phi-1-2} let us note that any of the pertinent auxiliary 3-index
antisymmetric tensor gauge fields, for instance, $A_{\m\n\l}$ can be in
particular parametrized in terms of 4 auxiliary scalar fields
$\{\phi^I\}_{I=1,\ldots, 4}$:
\be
A_{\m\n\l}=\frac{1}{4} \vareps_{IJKL}\,
\phi^I \pa_\m \phi^J \pa_\n \phi^K \pa_\l \phi^L \; ,
\lab{A-measure-scalars}
\ee
so that:
\be
\Phi_1 (A) = \frac{1}{4!} \vareps^{\m\n\k\l} \vareps_{IJKL}\,
\pa_\m \phi^I \pa_\n \phi^J \pa_\k \phi^K \pa_\l \phi^L 
= \det\Vert \frac{\pa \phi^I}{\pa x^\m} \Vert
\lab{Phi-measure-scalars}
\ee
acquires the form of a Jacobian. In a recent study \ct{struckmeier} of general
relativity as an extended canonical gauge theory a similar Jacobian
representation of the covariant integration measure has appeared in terms of
additional scalar fields. However, unlike the present case in the
construction of Ref.~\refcite{struckmeier} the additional scalar fields enter 
also in the proper Lagrangian.

In what follows we will stick to the representation \rf{Phi-1-2}.


The scalar field potentials and the separate locations of the standard Maxwell and the 
square-root Maxwell gauge field terms have been chosen in such a way that the
original action \rf{TMMT+GG} is invariant under global Weyl-scale transformations:
\br
g_{\m\n} \to \l g_{\m\n} \;\; ,\;\; \G^\m_{\n\l} \to \G^\m_{\n\l}
\;\; ,\;\;  \vp \to \vp + \frac{1}{\a}\ln \l \;\;, A_\m \to A_\m \; ,
\nonu \\
A_{\m\n\k} \to \l A_{\m\n\k} \;\; ,\;\; 
B_{\m\n\k} \to \l^2 B_{\m\n\k} \;\; ,\;\; H_{\m\n\k} \to H_{\m\n\k} \; .
\lab{scale-transf}
\er

The equations of motion resulting from the action \rf{TMMT+GG} are as follows.
Variation of \rf{TMMT+GG} w.r.t. affine connection $\G^\m_{\n\l}$:
\be
\int d^4\,x\,\sqrt{-g} g^{\m\n} \Bigl(\frac{\P_1}{\sqrt{-g}} +
2\eps\,\frac{\P_2}{\sqrt{-g}}\, R\Bigr) \(\nabla_\k \d\G^\k_{\m\n}
- \nabla_\m \d\G^\k_{\k\n}\) = 0 
\lab{var-G}
\ee
gives, following the analogous derivation in the Ref.~\refcite{TMT-orig-1}, that 
$\G^\m_{\n\l}$ becomes a Levi-Chevitta connection:
\be
\G^\m_{\n\l} = \G^\m_{\n\l}({\bar g}) = 
\h {\bar g}^{\m\k}\(\pa_\n {\bar g}_{\l\k} + \pa_\l {\bar g}_{\n\k} 
- \pa_\k {\bar g}_{\n\l}\) \; ,
\lab{G-eq}
\ee
w.r.t. to the Weyl-rescaled metric ${\bar g}_{\m\n}$:
\be
{\bar g}_{\m\n} = (\chi_1 + 2\eps \chi_2 R) g_{\m\n} \;\; ,\;\; 
\chi_1 \equiv \frac{\P_1 (A)}{\sqrt{-g}} \;\; ,\;\;
\chi_2 \equiv \frac{\P_2 (B)}{\sqrt{-g}} \; .
\lab{bar-g}
\ee

Variation of the action \rf{TMMT+GG} w.r.t. auxiliary tensor gauge fields
$A_{\m\n\l}$, $B_{\m\n\l}$ and $H_{\m\n\l}$ yields the equations:
\be
\pa_\m \Bigl\lb R + L^{(1)} \Bigr\rb = 0 \quad, \quad
\pa_\m \Bigl\lb L^{(2)} + \eps R^2 + \frac{\P (H)}{\sqrt{-g}}\Bigr\rb = 0 
\quad, \quad \pa_\m \Bigl(\frac{\P_2 (B)}{\sqrt{-g}}\Bigr) = 0 \; ,
\lab{A-B-H-eqs}
\ee
whose solutions read:
\be
\frac{\P_2 (B)}{\sqrt{-g}} \equiv \chi_2 = {\rm const} \;\; ,\;\;
R + L^{(1)} = - M_1 = {\rm const} \;\; ,\;\; 
L^{(2)} + \eps R^2 + \frac{\P (H)}{\sqrt{-g}} = - M_2  = {\rm const} \; .
\lab{integr-const}
\ee
Here $M_1$ and $M_2$ are arbitrary dimensionful and $\chi_2$
arbitrary dimensionless integration constants. 
The appearance of $M_1,\, M_2$ signifies {\em dynamical spontaneous
breakdown} of global Weyl-scale invariance under \rf{scale-transf} due to the 
scale non-invariant solutions (second and third ones) in \rf{integr-const}.

It is also very instructive to elucidate the physical meaning of the three arbitrary 
integration constants $M_1,\, M_2,\,\chi_2$ from the point of view of the
canonical Hamiltonian formalism. Namely, as shown in the \textsl{Appendix A} 
$M_1,\, M_2,\,\chi_2$ are identified as conserved Dirac-constrained
canonical momenta conjugated to (certain components of) the auxiliary
maximal rank antisymmetric tensor gauge fields $A_{\m\n\l}, B_{\m\n\l}, H _{\m\n\l}$
entering the original non-Riemannian volume-form action \rf{TMMT+GG}.

Varying \rf{TMMT+GG} w.r.t. $g_{\m\n}$ and using relations \rf{integr-const} 
we have:
\be
\chi_1 \Bigl\lb R_{\m\n} + \h\( g_{\m\n}L^{(1)} - T^{(1)}_{\m\n}\)\Bigr\rb -
\h \chi_2 \Bigl\lb T^{(2)}_{\m\n} + g_{\m\n} \(\eps R^2 + M_2\)
- 2 R\,R_{\m\n}\Bigr\rb = 0 \; ,
\lab{pre-einstein-eqs}
\ee
where $\chi_1$ and $\chi_2$ are defined in \rf{bar-g},
and $T^{(1,2)}_{\m\n}$ are the energy-momentum tensors of the scalar+gauge
field Lagrangians with the standard definitions:
\be
T^{(1,2)}_{\m\n} = g_{\m\n} L^{(1,2)} - 2 \partder{}{g^{\m\n}} L^{(1,2)} \; .
\lab{EM-tensor}
\ee

Taking the trace of Eqs.\rf{pre-einstein-eqs} and using again second relation 
\rf{integr-const} we solve for the scale factor $\chi_1$:
\be
\chi_1 = 2 \chi_2 \frac{T^{(2)}/4 + M_2}{L^{(1)} - T^{(1)}/2 - M_1} \; ,
\lab{chi-1}
\ee
where $T^{(1,2)} = g^{\m\n} T^{(1,2)}_{\m\n}$. 

Using second relation \rf{integr-const} Eqs.\rf{pre-einstein-eqs} can be put 
in the Einstein-like form:
\br
R_{\m\n} - \h g_{\m\n}R = \h g_{\m\n}\(L^{(1)} + M_1\)
+ \frac{1}{2\O}\(T^{(1)}_{\m\n} - g_{\m\n}L^{(1)}\)
\nonu \\
+ \frac{\chi_2}{2\chi_1 \O} \Bigl\lb T^{(2)}_{\m\n} + 
g_{\m\n} \(M_2 + \eps(L^{(1)} + M_1)^2\)\Bigr\rb \; ,
\lab{einstein-like-eqs}
\er
where:
\be
\O = 1 - \frac{\chi_2}{\chi_1}\,2\eps\(L^{(1)} + M_1\) \; .
\lab{Omega-eq}
\ee
Let us note that \rf{bar-g}, upon taking into account second relation
\rf{integr-const} and \rf{Omega-eq}, can be written as:
\be
{\bar g}_{\m\n} = \chi_1\O\,g_{\m\n} \; .
\lab{bar-g-2}
\ee

Now, we can bring Eqs.\rf{einstein-like-eqs} into the standard form of Einstein 
equations for the rescaled  metric ${\bar g}_{\m\n}$ \rf{bar-g-2}, 
\textsl{i.e.}, the Einstein-frame equations: 
\be
R_{\m\n}({\bar g}) - \h {\bar g}_{\m\n} R({\bar g}) = \h T^{\rm eff}_{\m\n}
\lab{eff-einstein-eqs}
\ee
with energy-momentum tensor corresponding according to the definition \rf{EM-tensor}:
\be
T^{\rm eff}_{\m\n} = g_{\m\n} L_{\rm eff} - 2 \partder{}{g^{\m\n}} L_{\rm eff}
\lab{T-eff}
\ee
to the following effective (Einstein-frame) scalar field Lagrangian:
\be
L_{\rm eff} = \frac{1}{\chi_1\O}\Bigl\{ L^{(1)} + M_1 +
\frac{\chi_2}{\chi_1\O}\Bigl\lb L^{(2)} + M_1 + 
\eps (L^{(1)} + M_1)^2\Bigr\rb\Bigr\} \; .
\lab{L-eff}
\ee

In order to explicitly write $L_{\rm eff}$ in terms of the Einstein-frame
metric ${\bar g}_{\m\n}$ \rf{bar-g-2} we use the short-hand notation for the
scalar kinetic term:
\be
X \equiv - \h {\bar g}^{\m\n}\pa_\m \vp \pa_\n \vp
\lab{X-def}
\ee
and represent $L^{(1,2)}$ in the form:
\be
L^{(1)} = \chi_1\O\, X - V - \chi_1\O\,\frac{f_0}{2}\sqrt{-F^2({\bar g})}
\quad ,\quad 
L^{(2)} = \chi_1\O\,b e^{-\a\vp}X + U - \(\chi_1\O\)^2 \frac{1}{4e^2}F^2({\bar g}) \; ,
\lab{L-1-2-Omega}
\ee
with $V$ and $U$ as in \rf{L-1}-\rf{L-2}.

From Eqs.\rf{chi-1} and \rf{Omega-eq}, taking into account \rf{L-1-2-Omega}, 
we find:
\be
\frac{1}{\chi_1\O} = \frac{(V-M_1)}{2\chi_2\Bigl\lb U+M_2 + \eps (V-M_1)^2\Bigr\rb}
\,\Bigl\lb 1 - \chi_2 \Bigl(\frac{b e^{-\a\vp}}{V-M_1} - 2\eps\Bigr) X
-\eps\chi_2 f_0 \sqrt{-F^2({\bar g})} \Bigr\rb \; .
\lab{chi-Omega}
\ee
Upon substituting expression \rf{chi-Omega} into \rf{L-eff} we arrive at the
explicit form for the Einstein-frame matter Lagrangian:
\be
L_{\rm eff} = A(\vp) X + B (\vp) X^2 - U_{\rm eff}(\vp)
- \frac{F^2({\bar g})}{4e_{\rm eff}^2(\vp)}
- \frac{f_{\rm eff}(\vp)}{2}\sqrt{-F^2({\bar g})}
- \eps\chi_2 f_0 A(\vp) X \sqrt{-F^2({\bar g})} \; .
\lab{L-eff-GG}
\ee
The coefficient functions in \rf{L-eff-GG} read:
\be
A(\vp) = 1 - 4 U_{\rm eff}(\vp) \Bigl\lb \eps\chi_2  
- \frac{\chi_2 b e^{-\a\vp}}{2(V(\vp)-M_1)} \Bigr\rb \quad ,\quad
B(\vp) = \eps\chi_2 - 4 U_{\rm eff}(\vp) \Bigl\lb \eps\chi_2  
- \frac{\chi_2 b e^{-\a\vp}}{2(V(\vp)-M_1)} \Bigr\rb^2 \; ,
\lab{A-B-GG}
\ee
whereas the effective scalar field potential reads:
\be
U_{\rm eff} (\vp) \equiv 
\frac{(V - M_1)^2}{4\chi_2 \Bigl\lb U + M_2 + \eps (V - M_1)^2\Bigr\rb}
= \frac{\(f_1 e^{-\a\vp}-M_1\)^2}{4\chi_2\,\Bigl\lb 
f_2 e^{-2\a\vp} + M_2 + \eps (f_1 e^{-\a\vp}-M_1)^2\Bigr\rb} \; ,
\lab{U-eff}
\ee
where the explicit form of $V$ and $U$ \rf{L-1}-\rf{L-2} are inserted.
Further, the original gauge coupling constants are here replaced by $\vp$-dependent
effective coupling constants:
\br
f_{\rm eff}(\vp) = f_0 \Bigl( 1 - 4\eps\chi_2 U_{\rm eff}(\vp)\Bigr)
\nonu \\
= f_0 \frac{f_2 e^{-2\a\vp}+M_2}{f_2 e^{-2\a\vp}+M_2 +\eps (f_1 e^{-\a\vp}-M_1)^2} \; ,
\lab{f-eff} \\
\frac{1}{e_{\rm eff}^2 (\vp)} = \chi_2 \Bigl\lb \frac{1}{e^2} +
\eps\,f_0^2 \Bigl( 1-4\eps\chi_2 U_{\rm eff}(\vp)\Bigr)\Bigr\rb
\nonu \\
= \chi_2 \Bigl\lb \frac{1}{e^2} + \eps\, f_0^2 
\frac{f_2 e^{-2\a\vp}+M_2}{f_2 e^{-2\a\vp}+M_2 +\eps (f_1 e^{-\a\vp}-M_1)^2}\Bigr\rb \; .
\lab{e-eff}
\er
Let us recall that the dimensionless integration constant $\chi_2$ systematically 
appearing in most relations is the ratio of the original second non-Riemannian 
integration measure to the standard Riemannian one \rf{bar-g}.

We observe that even if we start with {\em no} standard Maxwell kinetic
term for the gauge field, \textsl{i.e.}, taking the limit $e^2 \to \infty$
in the original action \rf{TMMT+GG}-\rf{L-2}, we nevertheless obtain 
a {\em dynamically induced} Maxwell term in the Einstein-frame action 
\rf{L-eff-GG} with effective running charge according to \rf{e-eff}:
\be
\frac{1}{e_{\rm eff}^2 (\vp)} = \eps\chi_2 f_0^2 \frac{(f_2 e^{-2\a\vp}+M_2)}{ 
\bigl\lb f_2 e^{-2\a\vp}+M_2 +\eps (f_1 e^{-\a\vp}-M_1)^2\bigr\rb} \; .
\lab{e-eff-dynamic}
\ee
From \rf{e-eff-dynamic} we see that dynamical Maxwell term generation is a
cumulative effect of the simultaneous presence of the ``confining'' gauge
field term $\sqrt{-F^2}$ and the $R^2$ gravity term.

\section{Flat Regions of the Effective Scalar Potential and Nontrivial
``Vacuum'' Solutions}
The explicit expressions for the effective potential $U_{\rm eff} (\vp)$
\rf{U-eff}, the scalar (``k-essence'') kinetic terms' coefficient functions 
$A(\vp)$ and $B(\vp)$ \rf{A-B-GG} and the effective gauge coupling constants
\rf{f-eff}-\rf{e-eff} reveal the following crucial feature of the Einstein-frame matter 
Lagrangian \rf{L-eff-GG}-\rf{e-eff}: the presence of {\em two infinitely large
flat regions} -- one for large negative and one for large positive values of the
scalar field $\vp$, where all of the above are essentially constant w.r.t. $\vp$. 

Depending on the sign of the integration constant $M_1$ we obtain two types of 
shapes for the effective scalar potential $U_{\rm eff} (\vp)$ \rf{U-eff}
depicted on Fig.1. and Fig.2.

\begin{figure}
\begin{center}
\includegraphics{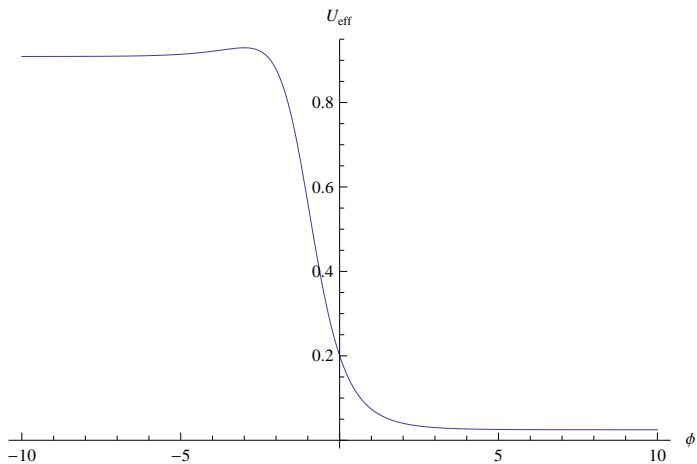}
\caption{Qualitative shape of the effective scalar potential $U_{\rm eff} (\vp)$ 
\rf{U-eff} for $M_1 > 0$.}
\end{center}
\end{figure}
\begin{figure}
\begin{center}
\includegraphics{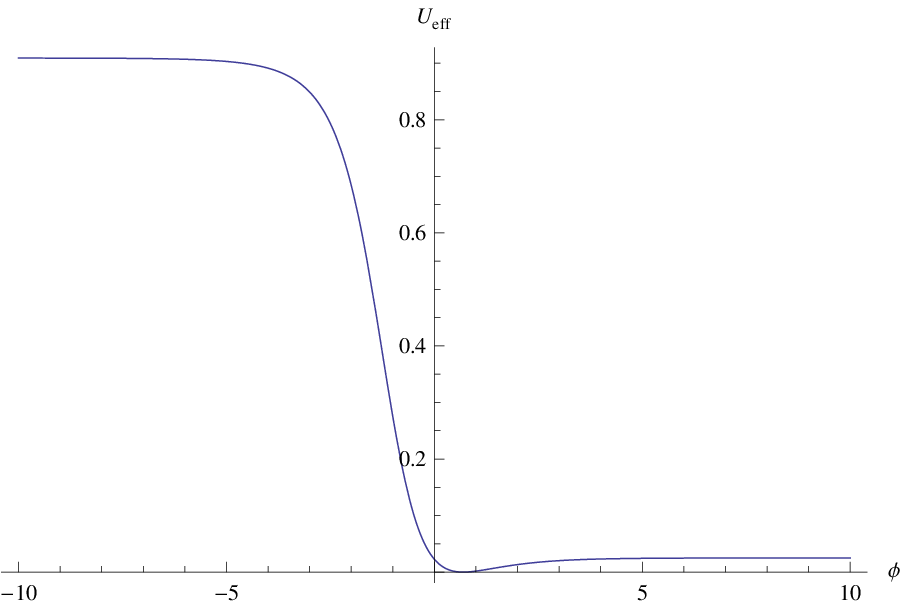}
\caption{Qualitative shape of the effective scalar potential $U_{\rm eff} (\vp)$ 
\rf{U-eff} for $M_1 < 0$.}
\end{center}
\end{figure}

For large negative values of $\vp$ we have for the effective potential and the
coefficient functions in the Einstein-frame matter Lagrangian 
\rf{L-eff-GG}-\rf{e-eff}:
\br
U_{\rm eff}(\vp) \simeq U_{(-)} \equiv 
\frac{f_1^2/f_2}{4\chi_2 (1+\eps f_1^2/f_2)} \; ,
\lab{U-minus} \\
A(\vp) \simeq A_{(-)} \equiv \frac{1+\h b f_1/f_2}{1+\eps f^2_1/f_2} \;\; ,\;\; 
B(\vp) \simeq B_{(-)} \equiv 
\eps\chi_2 \frac{1+ b f_1/f_2 - b^2/4\eps f_2}{1+\eps f^2_1/f_2} \; ,
\lab{A-B-minus} \\
e^2_{\rm eff} (\vp) \simeq e_{(-)} \equiv 
\frac{e^2}{\chi_2} \frac{1+\eps f_1^2/f_2}{1+\eps f_1^2/f_2 + e^2\eps f_0^2}
\;\; ,\;\;
f_{\rm eff}(\vp) \simeq f_{(-)} \equiv \frac{f_0}{1+\eps f_1^2/f_2} \; .
\lab{e-f-minus}
\er
This will be called ``(-) flat region''. In the second flat region for large
positive $\vp$, which will be called ``(+) flat region'' we have:
\br
U_{\rm eff}(\vp) \simeq U_{(+)} \equiv 
\frac{M_1^2/M_2}{4\chi_2 (1+\eps M_1^2/M_2)} \; ,
\lab{U-plus} \\
A(\vp) \simeq A_{(+)} \equiv \frac{M_2}{M_2 + \eps M_1^2} \quad ,\quad
B(\vp) \simeq B_{(+)} \equiv \eps\chi_2 \frac{M_2}{M_2 + \eps M_1^2} \; ,
\lab{A-B-plus} \\
e^2_{\rm eff} (\vp) \simeq e_{(+)} \equiv 
\frac{e^2}{\chi_2} \frac{1+\eps M_1^2/M_2}{1+\eps M_1^2/M_2 + e^2\eps f_0^2} 
\;\; ,\;\;
f_{\rm eff}(\vp) \simeq f_{(+)} \equiv \frac{f_0}{1+\eps M_1^2/M_2} \; .
\lab{e-f-plus}
\er

The scalar and gauge field equations of motion resulting from the
Einstein-frame Lagrangian (here $L_{\rm eff}$ is considered function of
$\vp, X, F^2$):
\be
\frac{1}{\sqrt{-{\bar g}}} \pa_\m \Bigl(\sqrt{-{\bar g}} {\bar g}^{\m\n} \pa_\n \vp
\partder{L_{\rm eff}}{X}\Bigr) - \partder{L_{\rm eff}}{\vp} = 0 \quad ,\quad
\pa_\n \Bigl(\sqrt{-{\bar g}} F^{\m\n} \partder{L_{\rm eff}}{F^2}\Bigr) = 0
\lab{L-eff-eqs}
\ee
thanks to the presence of the two $(\pm)$ large flat regions 
\rf{U-minus}-\rf{e-f-minus} and \rf{U-plus}-\rf{e-f-plus}, as well as due to the 
``k-essence''-type nonlinear dependence of $L_{\rm eff}$ on the scalar kinetic term,
allow for the following two classes of nontrivial ``vacuum'' solutions:

\begin{itemize}
\item
(i) ``Standard vacuum'' containing standard constant ``dilaton'' vacuum plus 
nontrivial gauge field vacuum:
\be
\vp = {\rm const} \;\; \to \;\; X = 0 \quad ,\quad
\partder{L_{\rm eff}}{\vp} = 0 \quad ,\quad \partder{L_{\rm eff}}{F^2} = 0 \; .
\lab{class-1}
\ee
Here the value $\vp = {\rm const}$ belongs to either the $(-)$ flat region
\rf{U-minus} or the $(+)$ flat region \rf{U-plus} of the effective scalar potential.
\item
(ii) ``Kinetic vacuum'' (this type of ``vacuum'' exists thanks to the nonlinear 
w.r.t. $X$ ``k-essence'' nature of the effective Lagrangian \rf{L-eff-GG}): 
\be
\partder{L_{\rm eff}}{X} = 0 \quad ,\quad
\partder{L_{\rm eff}}{\vp} = 0 \quad ,\quad \partder{L_{\rm eff}}{F^2} = 0 \; ,
\lab{class-2}
\ee
\end{itemize}

In the first class of ``standard vacuum'' solutions the last equation \rf{class-1}
yields the following non-trivial ``vacuum'' value for the gauge field:
\be
\sqrt{-F^2_{\rm vac}} = e^2_{\rm eff}(\vp) f_{\rm eff}(\vp) \; ,
\lab{F-vac-1}
\ee
and for the associated matter energy-momentum tensor (cf. \rf{T-eff}) we get:
\be
T^{\rm eff}_{\m\n} = 
{\bar g}_{\m\n} L_{\rm eff}\bv_{X=0,\,\partder{L_{\rm eff}}{F^2}=0} 
= - {\bar g}_{\m\n} U^{(\rm standard)}_{\rm total}
\lab{T-eff-standard}
\ee
where $U^{(\rm standard)}_{\rm total}$ is the total effective scalar potential
in the ``standard vacuum'' \rf{class-1}:
\be
U^{(\rm standard)}_{\rm total} = 
U_{\rm eff} + \frac{1}{4} e^2_{\rm eff} f^2_{\rm eff} =
U_{\rm eff} + \frac{e^2 f_0^2 \( 1-4\eps\chi_2 U_{\rm eff}\)^2}{
4\chi_2 \lb 1 + e^2\eps f_0^2 \( 1-4\eps\chi_2 U_{\rm eff}\)\rb}
\lab{U-standard}
\ee
In both $(\mp)$ flat regions \rf{U-minus}-\rf{e-f-minus} and \rf{U-plus}-\rf{e-f-plus}
we have correspondingly:
\br
\sqrt{-F^2_{\rm vac}} \simeq \sqrt{-F^2_{(-)}} =
\frac{e^2 f_0}{\chi_2\,(1+\eps f_1^2/f_2 + e^2\eps f_0^2)} \; ,
\lab{F-minus} \\
\sqrt{-F^2_{\rm vac}} \simeq \sqrt{-F^2_{(+)}} =
\frac{e^2 f_0}{\chi_2\,(1+\eps M_1^2/M_2 + e^2\eps f_0^2)} \; ,
\lab{F-plus}
\er
and:
\br
U^{(\rm standard)}_{\rm total} \simeq U^{(\rm standard)}_{(-)} \equiv
\frac{1}{4\eps\chi_2} \Bigl\lb 1 - 
\frac{1}{1+\eps f_1^2/f_2 + \eps e^2 f_0^2}\Bigr\rb \; ,
\lab{U-standard-minus} \\
U^{(\rm standard)}_{\rm total} \simeq U^{(\rm standard)}_{(+)} \equiv
\frac{1}{4\eps\chi_2} \Bigl\lb 1 - 
\frac{1}{1+\eps M_1^2/M_2 + \eps e^2 f_0^2}\Bigr\rb \; .
\lab{U-standard-plus}
\er
Therefore, according to \rf{T-eff-standard} the solutions of the Einstein-frame 
${\bar g}_{\m\n}$-equations are of de Sitter type (pure de Sitter or 
Schwarzschild-de Sitter):
\br
ds^2 = {\bar g}_{\m\n} dx^\m dx^\n = - \cA (r) dt^2 + \frac{dr^2}{\cA (r)}
+ r^2 \bigl( d\th^2 + \sin^2 \th d\phi\Bigr) \; ,
\lab{deSitter-type} \\
\cA (r) = 1 - \frac{\L_{(\pm)}}{3}r^2 \quad, \;\; {\rm or} \;\;
\cA (r) = 1 - \frac{2m}{r} - \frac{\L_{(\pm)}}{3}r^2 
\lab{dS-SdS-def}
\er
in static spherically symmetric coordinate chart,
with effective cosmological constants $\L_{(\pm)}$ given by 
\rf{U-standard-minus}-\rf{U-standard-plus}:
%
%
\br
\L_{(-)} \equiv \L^{(\rm standard)}_{(-)} = \h U^{(\rm standard)}_{(-)} \;\; 
{\rm in~the~}(-){\rm flat~region~\rf{U-minus}-\rf{e-f-minus}} \; ,
\lab{CC-standard-minus} \\
\L_{(+)} \equiv \L^{(\rm standard)}_{(+)} = \h U^{(\rm standard)}_{(+)} \;\; 
{\rm in~the~}(+){\rm flat~region~\rf{U-plus}-\rf{e-f-plus}} \; .
\lab{CC-standard-plus}
\er

From the above analysis of the ``standard vacuum'' solutions -- the one corresponding
to $\vp={\rm const}$ belonging to the $(-)$ flat region of the effective scalar potential
with nonzero gauge field vacuum value and vacuum energy density as in \rf{F-minus}, 
\rf{U-standard-minus}, and the second the one corresponding to $\vp={\rm const}$ 
belonging to the $(+)$ flat region of the effective scalar potential with gauge field
vacuum value and vacuum energy density as in \rf{F-plus}, \rf{U-standard-plus} -- 
we conclude that these ``standard vacuum''solutions describe {\em charge confining}
phases with dynamically generated cosmological constants \rf{CC-standard-minus}, 
\rf{CC-standard-plus}.

Indeed, according to `t Hooft's confinement proposal \ct{thooft-1,thooft-2}, and as
shown explicitly in Ref.~\refcite{GG-2} in the case of flat spacetime, 
the non-zero vacuum values of the gauge field 
\rf{F-minus}, \rf{F-plus} imply {\em confinement} dynamics of charged particles with 
the strength of confinement proportional to these same gauge field vacuum values.
Namely, under plausible truncation for static spherically symmetric configurations 
the canonically quantized theory in flat spacetime of charged fermions interacting with the 
nonlinear gauge fields with the ``square-root'' Maxwell term produces an
effective ``Cornell''-type potential $V_{\rm eff} = - \frac{\a}{L} + \b L$
(see also Eq.\rf{cornell} below) between quantized fermions separated by
a distance $L$ and where $\b$ is proportional to the coupling constant of the 
``square-root'' Maxwell term, which in turn is proportional to the non-zero
vacuum value of the gauge field.

The formalism to prove confinement used in Ref.~\refcite{GG-2} can be easily
generalized to the case of curved static spherically symmetric spacetimes,
in particular for de Sitter spacetime where both charged fermions are located within 
the interior de Sitter region below the de Sitter horizon ($r \leq \sqrt{3/\L_{(\pm)}}$) 
- see \textsl{Appendix B}.

\section{``Kinetic Vacuum'' Solutions}
\lab{kinetic-vac}

We now turn our attention to the second class of ``kinetic vacuum'' solutions 
\rf{class-2}. The equations $\partder{L_{\rm eff}}{X} = 0$ and
$\partder{L_{\rm eff}}{F^2} = 0$ yield:
\br
X_{\rm kin} = -\frac{A}{2B}\,\frac{1-\eps\chi_2 f_0 f_{\rm eff} e^2_{\rm eff}}{
1 - \eps^2\chi_2^2 f_0^2 e^2_{\rm eff} A^2/B} \; ,
\lab{X-vac-2} \\
\sqrt{-F^2_{\rm kin}} = e^2_{\rm eff} \frac{f_{\rm eff} - \eps\chi_2 f_0 A^2/B}{
1 - \eps^2\chi_2^2 f_0^2 e^2_{\rm eff} A^2/B} \; .
\lab{F-vac-2}
\er

Using the identity 
$\partder{L_{\rm eff}}{X} = 2BX + A(1-\eps\chi_2 f_0 \sqrt{-F^2})$ we can
rewrite the Einstein-frame Lagrangian $L_{\rm eff}$ \rf{L-eff} in the form:
\be
L_{\rm eff} = \frac{1}{4B}\Bigl(\partder{L_{\rm eff}}{X}\Bigr)^2 - {\wti U} (\vp, F^2)
\; ,
\lab{L-eff-new}
\ee
with:
\be
{\wti U} (\vp, F^2) = U_{\rm eff} + \frac{A^2}{4B}
- \h\sqrt{-F^2}\Bigl( f_{\rm eff}-\eps\chi_2 f_0 \frac{A^2}{B}\Bigr)
-\frac{1}{4}F^2 \Bigl(\frac{1}{e^2_{\rm eff}} -
\eps^2 \chi_2^2 f_0^2 \frac{A^2}{B}\Bigr) \; .
\lab{U-wti}
\ee
Inserting in \rf{L-eff-new}-\rf{U-wti} the ``on-shell'' values 
\rf{X-vac-2}-\rf{F-vac-2}, we obtain for the matter energy-momentum tensor:
\be
T^{\rm eff}_{\m\n} = 
{\bar g}_{\m\n} 
L_{\rm eff}\bv_{\partder{L_{\rm eff}}{X}=0,\,\partder{L_{\rm eff}}{F^2}=0} 
= - {\bar g}_{\m\n} U^{(\rm kinetic)}_{\rm total} \; ,
\lab{T-eff-kinetic}
\ee
where $U^{(\rm kinetic)}_{\rm total}$ is the total effective scalar potential
in the ``kinetic vacuum'' \rf{class-2}:
\be
U^{(\rm kinetic)}_{\rm total} = U_{\rm eff} + \frac{A^2}{4B} +
\frac{1}{4} e^2_{\rm eff}
\frac{\Bigl( f_{\rm eff}-\eps\chi_2 f_0 \frac{A^2}{B}\Bigr)^2}{
1 - e^2_{\rm eff}\eps^2 \chi_2^2 f_0^2 \frac{A^2}{B}} \; .
\lab{U-kinetic}
\ee

Note from Eqs.\rf{L-eff-new}-\rf{U-wti} that within the ``kinetic vacuum'' 
$\partder{L_{\rm eff}}{X}=0$ the effective gauge coupling constants become:
\be
{\wti f}_{\rm eff} = f_{\rm eff}-\eps\chi_2 f_0 \frac{A^2}{B} \quad, \quad
{\wti e}^2_{\rm eff} = 
\frac{e^2_{\rm eff}}{1 - e^2_{\rm eff}\eps^2 \chi_2^2 f_0^2 \frac{A^2}{B}}
\lab{e-f-kinetic-vac}
\ee

\subsection{``Kinetic Vacuum'' in the $(+)$ Flat Region of Effective Scalar
Potential}
\lab{kinetic-vac-plus}

First we consider the ``kinetic vacuum'' solution in the $(+)$ flat region. 
Using \rf{U-plus}-\rf{e-f-plus} in \rf{F-vac-2}, \rf{X-vac-2} and \rf{U-kinetic} 
we obtain from \rf{e-f-kinetic-vac}:
\be
{\wti f}_{\rm eff} \equiv {\wti f}_{(+)} = 
f_{(+)}-\eps\chi_2 f_0 \frac{A^2_{(+)}}{B_{(+)}} = 0 ,
\lab{f-kinetic-plus}
\ee
which yields:
\br
\sqrt{-F^2_{\rm kin}}\bv_{(+)} = 0 \quad ,\quad
X_{\rm kin} \simeq X_{(+)} = - \frac{A_{(+)}}{2 B_{(+)}} = - \frac{1}{2\eps\chi_2}
\lab{vac-kinetic-plus} \\
U^{(\rm kinetic)}_{\rm total} \simeq U^{(\rm kinetic)}_{(+)} =
\frac{1}{4\eps\chi_2} \quad \to \;\; 
T^{\rm eff}_{\m\n} = - {\bar g}_{\m\n}\frac{1}{4\eps\chi_2}\; ,
\lab{U-kinetic-plus}
\er
\textsl{i.e.}, we have here an effective cosmological constant:
\be
\L_{(+)} \equiv \L^{(\rm kinetic)}_{(+)} = \frac{1}{8\eps\chi_2} \; .
\lab{CC-eff-plus}
\ee

Let us particularly stress on the first relation in \rf{vac-kinetic-plus} -- the zero 
vacuum value for the nonlinear gauge field, which is due to the vanishing 
\rf{f-kinetic-plus} of the effective coupling constant of the ``square-root'' Maxwell term. 
Again, in accordance with 
`t Hooft's confinement proposal \ct{thooft-1,thooft-2} and as demonstrated explicitly 
in Ref.~\refcite{GG-2} and in \textsl{Appendix B} the latter implies absence of confinement
of charged particles, \textsl{i.e.}, the ``kinetic vacuum'' 
\rf{vac-kinetic-plus}-\rf{U-kinetic-plus} describes a {\em deconfinement} phase.

According to \rf{T-eff-kinetic} and \rf{U-kinetic-plus}-\rf{CC-eff-plus}
the solutions of the Einstein-frame ${\bar g}_{\m\n}$-equations in the
``kinetic vacuum'' are again of de Sitter type \rf{deSitter-type}-\rf{dS-SdS-def}
with $\L_{(+)}$ given by \rf{CC-eff-plus}.

The equation for the ``dilaton'' ``kinetic vacuum'' (second Eq.\rf{vac-kinetic-plus})
reads explicitly:
\be
{\bar g}^{\m\n} \pa_\m \vp \pa_\n \vp - \frac{1}{\eps\chi_2} = 0 \; .
\lab{HJ-eq}
\ee
It has precisely the form of Hamilton-Jacobi equation for the
Hamilton-Jacobi action: 
\be
S(x) \equiv \vp (x) = \frac{1}{\sqrt{\eps\chi_2}} \int_{\l_{\rm in}}^{\l_{\rm out}}
d\l \sqrt{g_{\m\n} (x(\l)) \frac{d x^\m}{d\l} \frac{d x^\n}{d\l}}
\lab{HJ-action}
\ee
corresponding to spacelike geodesics $x^\m (\l)$ starting from some fixed
point $x_{(0)}$ (\textsl{e.g.}, $x_{(0)} = 0$) at a fixed value of the affine parameter
$\l_{\rm in}$ and passing through $x = x(\l_{\rm out})$ at $\l_{\rm out}$. 
This Hamilton-Jacobi action \rf{HJ-action} 
measures the proper distance between the points $x_{(0)}$ and $x$ on the
manifold modulo the numerical factor $1/\sqrt{\eps\chi_2}$ (for modern pedagogical exposition,
see, \textsl{e.g.}, Ref.~\refcite{a-hamilton}).

A static spherically symmetric solution for $\vp (x)$ is given by:
\be
\Bigl(\frac{\pa \vp}{\pa r}\Bigr)^2 = \frac{1}{\eps\chi_2\,\cA (r)} \quad \to
\;\; \vp (r) = \vp_{(+)} + 
\frac{1}{\sqrt{\eps\chi_2}}\int^r \frac{dr^\pr}{\sqrt{\cA (r^\pr)}} \; ,
\lab{vp-vac-plus}
\ee
where the initial value $\vp_{(+)}$ must belong to the $(+)$ flat region
\rf{U-plus}-\rf{e-f-plus} (large positive $\vp$).

In the case of pure de Sitter metric \rf{deSitter-type}-\rf{dS-SdS-def} 
the solution $\vp (r)$ \rf{vp-vac-plus}, measuring the proper radial distance
between $0$ and $r$, is clearly defined only for $r$ in the interval $r \in (0,r_{(+)})$, 
where:
\be
r_{(+)} = \sqrt{24\eps\chi_2}
\lab{r-plus}
\ee
is the de Sitter horizon radius. The solution $\vp (r)$ reads explicitly:
\be
\vp (r) = \vp_{(+)} + \sqrt{24}\arcsin\bigl(\frac{r}{r_{(+)}}\bigr) \; ,
\lab{vp-vac-plus-dS}
\ee
where the initial value $\vp_{(+)}$ belongs to the $(+)$ flat region of the
effective scalar potential.
Let us also recall that the integral in the second Eq.\rf{vp-vac-plus},
which is equal to $r_{(+)} \arcsin\bigl(\frac{r}{r_{(+)}}\bigr)$,
yields the proper radial distance in the internal de Sitter region $r\leq r_{(+)}$.

In the case of Schwarzschild-de Sitter metric \rf{deSitter-type}-\rf{dS-SdS-def}
the solution $\vp (r)$ \rf{vp-vac-plus} is defined in the interval
$r \in (r_S,r_H)$ between the inner (Schwarzschild-type) horizon $r_S$ and the outer
(de Sitter-type) horizon $r_H$. In what follows we will consider the case of
pure de Sitter metric.

Since the ``kinetic vacuum'' corresponding to the $(+)$ flat region described 
by \rf{vac-kinetic-plus}-\rf{vp-vac-plus-dS} is defined only within the
finite-volume space region below the de Sitter horizon, in order to be extended 
to the whole space it must be matched to another spherically symmetric 
configuration with the standard constant ``dilaton'' vacuum defined in the outer region 
beyond the de Sitter horizon with:
\be
\vp = \vp(r_{(+)}) = \vp_{(+)} + \sqrt{6}\pi = {\rm const} \;\;
{\rm for}\; r > r_{(+)} \; ,
\lab{vp-match-plus}
\ee
where the latter is the limiting value of \rf{vp-vac-plus-dS} at the horizon.
The corresponding construction yields a gravitational bag-like solution mimicking 
both some of the features of the MIT bags in QCD phenomenology \ct{MIT-bag-1,MIT-bag-2}
as well as some of the features of the ``constituent quark'' model of 
Ref.~\refcite{const-quark} to be discussed in the next Section.

\subsection{``Kinetic Vacuum'' in the $(-)$ Flat Region of Effective Scalar
Potential}
\lab{kinetic-vac-minus}

Next, let us consider the ``kinetic vacuum'' solution corresponding to the
$(-)$ flat region. In this case using \rf{U-minus}-\rf{e-f-minus} in 
\rf{F-vac-2}, \rf{X-vac-2} and \rf{U-kinetic} and introducing short-hand notations 
for some combinations of the parameters to simplify the resulting expressions:
\be
\xi \equiv b \frac{f_1}{f_2} \quad ,\quad \g \equiv \frac{f_2}{4\eps f_1^2}
\quad ,\quad \b \equiv e^2 \frac{f_0^2 f_2^2}{16\eps f_1^4} 
= \eps e^2 f_0^2 \g^2\; ,
\lab{param-minus-flat}
\ee
we obtain:
\br
\sqrt{-F^2_{\rm kin}} \simeq \sqrt{-F^2_{\rm kin}}\bv_{(-)} = \frac{\b \xi^4}{\eps\chi_2
(1+\xi - \xi^2 \g) \lb 1 +\xi - \xi^2 \g (1+ \b/\g^2)\rb} \; ,
\lab{F-kinetic-minus} \\
X_{\rm kin} \simeq X_{(-)} = - \frac{1}{2\eps\chi_2}\, 
\frac{1+\xi/2}{1 +\xi - \xi^2 \g (1+ \b/\g^2)} \; ,
\lab{X-kinetic-minus} \\
U^{(\rm kinetic)}_{\rm total} \simeq U^{(\rm kinetic)}_{(-)} = 
\frac{1}{4\eps\chi_2}\, 
\frac{1 + \xi - \xi^2 \b/\g}{1 +\xi - \xi^2 \g (1+ \b/\g^2)} \; ,
\lab{U-kinetic-minus} \\
T^{\rm eff}_{\m\n} = - {\bar g}_{\m\n} U^{(\rm kinetic)}_{(-)}
\equiv - {\bar g}_{\m\n} 2\L_{(-)} \; .
\lab{T-eff-plus}
\er
The spacetime metric is again of de Sitter type \rf{deSitter-type}-\rf{dS-SdS-def}
with effective cosmological constant:
\be
\L_{(-)} \equiv \L^{(\rm kinetic)}_{(-)} =
\frac{1}{8\eps\chi_2}\, 
\frac{1 + \xi - \xi^2 \b/\g}{1 +\xi - \xi^2 \g (1+ \b/\g^2)} \; .
\lab{CC-eff-minus}
\ee

According to Eq.\rf{F-kinetic-minus} the vacuum value of the gauge field is
non-zero (except in the special case $\xi=0$, see below \rf{simplify-minus}), 
therefore, following `t Hooft's confinement proposal
\ct{thooft-1,thooft-2} and Refs.~\refcite{GG-1}-\refcite{GG-6} we conclude that the 
``kinetic vacuum'' \rf{F-kinetic-minus}-\rf{U-kinetic-minus} supports dynamics of 
charged particles as in the ``standard vacuum'' phase \rf{class-1} except in
the special case $\xi=0$.

The qualitative shape of the ``kinetic vacuum'' energy density in the $(-)$ flat region
$U^{(\rm kinetic)}_{(-)} (\xi)$ \rf{U-kinetic-minus} as function of the parameter 
$\xi \equiv b \frac{f_1}{f_2}$ \rf{param-minus-flat} is depicted on Fig.3.

\begin{figure}
\begin{center}
\includegraphics{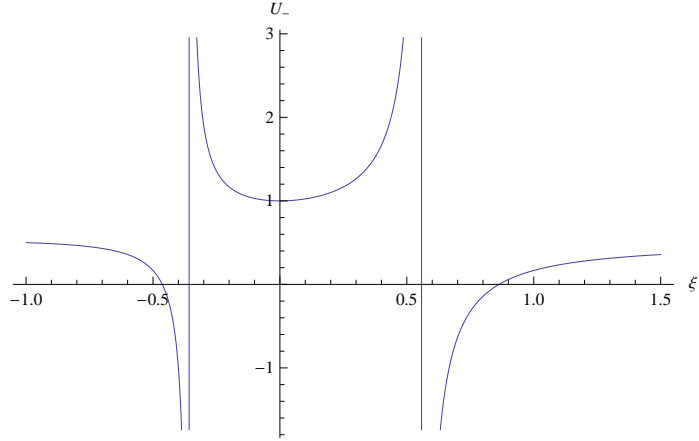}
\caption{Qualitative shape of the ``kinetic vacuum'' energy density \rf{U-kinetic-minus}
as function of $\xi \equiv b \frac{f_1}{f_2}$ \rf{param-minus-flat}.}
\end{center}
\end{figure}

$U^{(\rm kinetic)}_{(-)} (\xi)$  has a local minimum at $\xi=0$, \textsl{i.e.}, 
at $b=0$ where:
\be
U^{(\rm kinetic)}_{(-)} (0) = 1/4\eps\chi_2 \; ,
\lab{U-kinetic-minus-min}
\ee
and it raises to $+\infty$ at $\xi = \xi_\g^{(\pm)}$ which are the roots of the 
quadratic expression $1 +\xi - \xi^2 \g (1+ \b/\g^2$:
\be
\xi_\g^{(\pm)} = 
\frac{1}{2\g (1+\b/\g^2)}\Bigl\lb 1 \pm \sqrt{1+4\g (1+\b/\g^2)}\Bigr\rb \; .
\lab{roots-plus-minus}
\ee
Alternatively, for large $|\xi|$, $U^{(\rm kinetic)}_{(-)} (\xi\to\pm\infty) =
\frac{1}{4\eps\chi_2}\bigl(1+1/\eps e^2 f_0^2\Bigr)^{-1}$.

The scalar field for static spherically symmetric configurations is given by:
\be
\Bigl(\frac{\pa \vp}{\pa r}\Bigr)^2 = - 2\frac{X_{(-)}}{\cA (r)} \; ,
\lab{vp-eq-minus}
\ee
therefore, solutions exist only for the range of parameters for which $X_{(-)} <0$.
From the explicit expression \rf{X-kinetic-minus} we find:
\be
X_{(-)} <0 \;\; {\rm for}\;\; \xi_\g^{(-)} < \xi < \xi_\g^{(+)} \;\; 
{\rm and ~for}\;\; \xi < -2 \; ,
\lab{X-minus-range}
\ee
where $\xi_\g^{(\pm)}$ are the same as in \rf{roots-plus-minus}. Thus, the solution 
of \rf{vp-eq-minus} is similar to \rf{vp-vac-plus-dS}:
\be
\vp (r) = \vp_{(-)} + \sqrt{2|X_{(-)}|} \int^r \frac{dr^\pr}{\sqrt{\cA (r^\pr)}} = 
\vp_{(-)} + \sqrt{2|X_{(-)}|} r_{(-)} \arcsin\bigl(r/r_{(-)}\bigr) \; ,
\lab{vp-vac-minus}
\ee
where the initial value $\vp_{(-)}$ must belong to the $(-)$ flat region
\rf{U-minus}-\rf{e-f-minus} (large negative $\vp$), $r_{(-)}$ is the de Sitter 
horizon radius:
\be
r_{(-)} = \sqrt{3/\L_{(-)}} \quad (\L_{(-)} {\rm as ~in ~\rf{CC-eff-minus}}) \; ,
\lab{r-minus}
\ee
$|X_{(-)}|$ is given by \rf{X-kinetic-minus}, 
and again $\vp (r)$ is defined only in the space region inside the de Sitter
horizon ($r \leq r_{(-)}$).

Thus, similarly to the previous case for the $(+)$ flat region, here in the
$(-)$ flat region again we
need to match the ``kinetic vacuum'' given by \rf{F-kinetic-minus}-\rf{vp-vac-minus} 
to another spherically symmetric configuration with the standard constant dilaton
vacuum:
\be
\vp = \vp(r_{(-)}) = \vp_{(-)} + \pi r_{(-)}\sqrt{|X_{(-)}|/2}={\rm const}\;\;
{\rm for}\;\; r > r_{(-)} \; ,
\lab{vp-match-minus}
\ee
defined in the outer region, where $\vp(r_{(-)})$ is the limiting value of $\vp(r)$ 
\rf{vp-vac-minus} at the horizon. This will be considered in the next Section.

Let us specifically note that in the special case $\xi \equiv b \frac{f_1}{f_2}=0$,
\textsl{i.e.}, for $b=0$ meaning that in this case the non-canonical scalar kinetic
term is absent from the original second Lagrangian $L^{(2)}$ \rf{L-2}, 
the expressions \rf{F-kinetic-minus}-\rf{U-kinetic-minus} and \rf{CC-eff-minus}, 
\rf{r-minus} drastically simplify:
\br
\Bigl(\sqrt{-F^2_{\rm kin}}\bv_{(-)}\Bigr)\bv_{\xi=0} = 0 \quad ,\quad 
X_{(-)}\bv_{\xi=0} = - \frac{1}{2\eps\chi_2} \quad ,\quad 
U^{(\rm kinetic)}_{(-)}\bv_{\xi=0} = \frac{1}{4\eps\chi_2} \; ,
\nonu \\
\L^{(\rm kinetic)}_{(-)}\bv_{\xi=0} = \frac{1}{8\eps\chi_2} \quad ,\quad
r_{(-)}\bv_{\xi=0} = \sqrt{24\eps\chi_2} \; .
\lab{simplify-minus}
\er 
Expressions \rf{simplify-minus} precisely coincide with the corresponding values of 
$\sqrt{-F^2_{\rm kin}}\bv_{(+)}=0$, $X_{(+)},\, U^{(\rm kinetic)}_{(+)},\, 
\L^{(\rm kinetic)}_{(+)},\, r_{(+)}$
\rf{vac-kinetic-plus}-\rf{CC-eff-plus}, \rf{r-plus} in the $(+)$ flat region 
of the effective scalar potential. In particular, from the first relation in 
\rf{simplify-minus} we conclude that in the special case $\xi=0$ ($b=0$) the
corresponding ``kinetic vacuum'' on the $(-)$ flat region of the effective
scalar potential implies deconfinement in complete analogy with the ``kinetic vacuum'' 
on the $(+)$ flat region (subsection 4.1).

\section{Gravitational Bag-like Solutions}

\subsection{Matching ``Kinetic Vacuum'' to Standard Constant ``Dilaton''
Vacuum in $(+)$ flat region}

First, we construct matching of the ``kinetic vacuum'' in $(+)$ flat region
of the effective scalar potential given by de Sitter metric 
\rf{deSitter-type}-\rf{dS-SdS-def} in the interior region ($r < r_{(+)}$)
below the de Sitter horizon $r_{(+)}=\sqrt{24\eps\chi_2}$
with effective cosmological constant \rf{CC-eff-plus} and by 
Eqs.\rf{vac-kinetic-plus}-\rf{vp-vac-plus-dS}, to a static spherically symmetric
configuration containing the standard constant ``dilaton'' vacuum 
\rf{vp-match-plus} in the outer region ($r > r_{(+)}$) beyond the de Sitter horizon. 
The ``matching'' specifically means that the ``dilaton'' field, the gauge field
strength and the metric with its first derivatives must be continuous across
the horizon, in particular, the de Sitter horizon of the interior metric 
must coincide with a horizon of the exterior metric.

Obviously, the static spherically symmetric configuration in the outer region 
cannot be the ``standard vacuum'' of the full ``dilaton'' plus gauge field subsystem 
given by \rf{class-1}, \rf{F-vac-1} and \rf{U-standard-plus}, since:

(a) In the inner ``kinetic vacuum'' region $\sqrt{-F^2_{\rm kin}}\bv_{(+)} = 0$ 
(first Eq.\rf{vac-kinetic-plus}), whereas in the outer ``standard vacuum'' region
$\sqrt{-F^2_{(+)}} \neq 0$ \rf{F-plus},
which would imply that there should be a lightlike (``null'') brane with a non-zero 
surface electric charge density located on de Sitter horizon to account for
the jump of the gauge field strength across the horizon;

(b) The effective cosmological constant in the outer ``standard vacuum'' region
\rf{CC-standard-plus} is different and smaller than the effective cosmological constant 
\rf{CC-eff-plus} in the inner ``kinetic vacuum'' region.

In Refs.~\refcite{f(R)-maxwell,Beograd-2012} (see also the earlier works 
\ct{our-sqroot-1}${}^{-}$\ct{our-sqroot-3})
we have already explicitly derived static spherically symmetric
solutions of the coupled gravity/nonlinear gauge field/scalar ``dilaton'' system
\rf{L-eff-GG} with a generalized Reissner-Nordstr\"{o}m-(anti)de Sitter geometry
carrying a non-vanishing background constant radial electric field in addition to 
the standard Coulomb field. We will use this type of solution in the outer region 
beyond the de Sitter horizon to be matched with the ``kinetic vacuum'' 
\rf{vac-kinetic-plus}-\rf{vp-vac-plus-dS} in the interior region.

Specifically, for $r > r_{(+)}=\sqrt{24\eps\chi_2}$ the solution reads 
\ct{f(R)-maxwell,Beograd-2012} (the additional factor $\frac{1}{16\pi}$ in
Eq.\rf{metric-out-2} below is due to the adopted normalization for the
Newton constant $G_{\rm Newton}= 1/16\pi$):
\br
ds^2 = - \cA_{\rm out} (r) dt^2 + \frac{dr^2}{\cA_{\rm out} (r)}
+ r^2 \bigl( d\th^2 + \sin^2 \th d\phi\Bigr) \; ,
\lab{metric-out-1} \\
\cA_{\rm out} (r) = 1 + \frac{1}{16\pi} \Bigl\lb - \sqrt{8\pi}|Q| f_{(+)} - \frac{2m}{r}
+\frac{Q^2}{e^2_{(+)}\,r^2}\Bigr\rb - \frac{\L_{\rm out}}{3}r^2 \; , 
\lab{metric-out-2} \\
\L_{\rm out} = \h U^{(\rm standard)}_{(+)} = 
\frac{1}{8\eps\chi_2} \Bigl\lb 1 - 
\frac{1}{1+\eps M_1^2/M_2 + \eps e^2 f_0^2}\Bigr\rb
\;\; ({\rm as ~in ~\rf{U-standard-plus}, \rf{CC-standard-plus}}) \; ,
\nonu \\
\sqrt{-F^2_{\rm out}}(r) = \sqrt{2} |\vec{E}_{\rm out}(r)|= 
e^2_{(+)} f_{(+)} - \frac{|Q|}{\sqrt{2\pi}\,r^2} \quad ,\quad
(e^2_{(+)}\, ,\, f_{(+)} \;  {\rm as ~in ~\rf{e-f-plus}}) \; ,
\lab{F-out} \\
\vp = \vp(r_{(+)}) = {\rm const} \;\; ({\rm as ~in ~\rf{vp-match-plus}}) \; .
\lab{vp-out}
\er
Notice that in \rf{F-out} we have taken the constant radial background
electric field and the Coulomb field with opposite directions. Let us also
point out that the scalar potential corresponding to the static radial electric
field in \rf{F-out}:
\be
E^r_{\rm out} = - F_{0r} = \pa_r A_0 (r) \quad ,\quad 
A_0 (r) = \frac{1}{\sqrt{2}} e^2_{(+)} f_{(+)}\, r + \frac{|Q|}{\sqrt{4\pi}\,r} \; ,
\lab{cornell}
\ee
which is a static spherically symmetric solution of the nonlinear gauge field 
equations (last Eq.\rf{L-eff-eqs} with $L_{\rm eff}$ as in \rf{L-eff-GG} and with 
$X=0$ -- ``standard dilaton vacuum''), resembles the form of the well-known 
phenomenological ``Cornell potential'' in QCD which contains both a linear 
confining and a standard Coulomb part \ct{cornell-1}${}^{-}$\ct{cornell-3}.

The mass and electric charge parameters $(m,Q)$ in \rf{metric-out-2} are to be 
determined from the matching at the common horizon:
\br
\cA_{\rm out} (r_{(+)}) = \cA_{\rm in} (r_{(+)}) = 0 \quad ,\quad
\pa_r \cA_{\rm out} (r_{(+)}) = \pa_r \cA_{\rm in} (r_{(+)}) \; ,
\lab{metric-match} \\
\sqrt{-F^2_{\rm out}}(r_{(+)}) = \sqrt{-F^2_{\rm kin}}\bv_{(+)} = 0 \;\; 
({\rm according ~to ~first ~Eq.\rf{vac-kinetic-plus}}) \; , 
\lab{F-match}
\er
with the ``kinetic vacuum'' \rf{vac-kinetic-plus}-\rf{vp-vac-plus-dS}
for $r < r_{(+)}=\sqrt{24\eps\chi_2}$ with:
\br
ds^2 = - \cA_{\rm in} (r) dt^2 + \frac{dr^2}{\cA_{\rm in} (r)}
+ r^2 \bigl( d\th^2 + \sin^2 \th d\phi\Bigr) \; ,
\lab{metric-in-1} \\
\cA_{\rm in } (r) = 1 - \frac{\L_{\rm in}}{3}r^2 \quad ,\quad
\L_{\rm in} = \h U^{(\rm kinetic)}_{(+)} = 1/8\eps\chi_2 \; .
\lab{metric-in-2}
\er

Inserting \rf{F-out} into \rf{F-match} we determine $Q$:
\be
|Q| = \sqrt{2\pi} e^2_{(+)} f_{(+)}\, 24\eps\chi_2 
= \frac{\sqrt{2\pi}24\eps e^2 f_0 }{1+\eps M_1^2/M_2 + e^2\eps f_0^2} \; .
\lab{Q-sol}
\ee
Now, inserting \rf{Q-sol} into \rf{metric-match} yields:
\be
m=0 \; ,
\lab{m-sol}
\ee
and the following relation between the integration constants $M_{1,2}$ and
the initial coupling constants $\eps, e, f_0$:
\be
1+\eps \frac{M_1^2}{M_2} - 3 \eps f_0^2 e^2 = 0 \; .
\lab{coupling-rel}
\ee

To recapitulate, we have obtained the following ``vacuum-like'' solution:

\begin{itemize}
\item
In the inner space region $r <  r_{(+)}=\sqrt{24\eps\chi_2}$ we have an interior
de Sitter region  \rf{metric-in-1}-\rf{metric-in-2} below the de Sitter horizon at 
$r=r_{(+)}$ with effective cosmological constant \rf{metric-in-2}, with vanishing
vacuum gauge field (first Eq.\rf{vac-kinetic-plus}), ``kinetic vacuum''  scalar 
``dilaton'' according to \rf{vp-vac-plus-dS} and vacuum energy density 
\rf{U-kinetic-plus}:
\be
\rho_{\rm in} \simeq U^{(\rm kinetic)}_{(+)} = \frac{1}{4\eps\chi_2} \; .
\lab{rho-in}
\ee
\item
In the outer space region $r >  r_{(+)}=\sqrt{24\eps\chi_2}$ we have static
spherically symmetric metric \rf{metric-out-2} with:
\be
\cA_{\rm out}(r) =  1 - \frac{1}{2\eps e^2 f_0^2} + 
\frac{6\chi_2}{e^2 f_0^2}\,\frac{1}{r^2} 
- \frac{r^2}{24\eps\chi_2} \Bigl( 1 - \frac{1}{4\eps e^2 f_0^2}\Bigr) \; ,
\lab{metric-out-3}
\ee
where we have used the explicit expressions for $e^2_{(+)}, f_{(+)}$ \rf{e-f-plus},
$Q$ \rf{Q-sol} and $U^{(\rm standard)}_{(+)}$ \rf{U-standard-plus} together with 
the relation \rf{coupling-rel}.
The metric with \rf{metric-out-3} has de Sitter type horizon again at $r=r_{(+)}$
where the relation \rf{coupling-rel} among the parameters holds. 
\item
The outside nonlinear gauge field \rf{F-out} is a static radial electric field of
the explicit form:
\be
\sqrt{-F^2_{\rm out}}(r) = \sqrt{2}\, |E^r_{\rm out} (r)| = 
\frac{1}{4\eps\chi_2 f_0} \Bigl( 1 - \frac{24\eps\chi_2}{r^2} \Bigr) \; ,
\lab{F-out-r}
\ee
where again we have used \rf{e-f-plus} and \rf{coupling-rel}.
In \rf{F-out-r} there is a Coulomb piece in addition to a non-zero background constant 
radial electric field:
\be
|E^r_{\rm background}| = \frac{1}{\sqrt{2}} e^2_{(+)} f_{(+)} = 
\frac{1}{\sqrt{2}\,4\eps\chi_2 f_0} \; .
\lab{background-el-field}
\ee
Thanks to the latter the Coulomb field is completely cancelled at the horizon.
\item
The scalar ``dilaton'' is constant \rf{vp-out} and the energy density 
($\rho = - T^0_0$) reads (using again \rf{e-f-plus}, \rf{U-standard-plus} 
and \rf{coupling-rel}):
\br
\rho_{\rm out} (r) \simeq U^{(\rm standard)}_{(+)} - e^2_{(+)} f^2_{(+)}
\Bigl(\frac{r^2_{(+)}}{2r^2} - \frac{r^4_{(+)}}{4r^4}\Bigr) 
\nonu \\
= \frac{1}{4\eps\chi_2} \Bigl( 1 - \frac{1}{4\eps e^2 f_0^2}\Bigr)
- \frac{1}{\eps e^2 f_0^2\, r^2} \Bigl( 1 - \frac{24\eps\chi_2}{r^2} \Bigr) \; .
\lab{rho-out-1}
\er
Obviously (recall in \rf{rho-out-1} $r > r_{(+)} \equiv \sqrt{24\eps\chi_2}$):
\be
\rho_{\rm out} (r) \leq U^{(\rm standard)}_{(+)} = 
\frac{1}{4\eps\chi_2} \Bigl\lb 1 - 
\frac{1}{1+\eps M_1^2/M_2 + \eps e^2 f_0^2}\Bigr\rb 
< \rho_{\rm in} = \frac{1}{4\eps\chi_2} \; .
\lab{rho-out-in}
\ee
\end{itemize}

The above solution \rf{metric-out-1}-\rf{rho-out-in} is a gravitational bag-like 
configuration on the $(+)$ flat region of the effective scalar potential 
which mimics some of the properties of the MIT bag \ct{MIT-bag-1,MIT-bag-2}. 
Indeed, as already noticed in Section 3:


(i) In the inner finite volume space region below the horizon ($r < r_{(+)}$) the
vanishing vacuum value of the gauge field (first Eq.\rf{vac-kinetic-plus}) implies 
absence of confinement of charged particles \ct{GG-1}${}^{-}$\ct{GG-6}.

(ii) According to \rf{rho-out-in} the vacuum energy density $\rho_{\rm in}$ in the
inner finite volume space region (for $r < r_{(+)}$) is larger than the energy 
density $\rho_{\rm out}$ in the outside region.

There are, however, other properties of the present gravitational ``bag''
solution which are substantially different from those of the MIT bag and
which rather resemble some of the properties of the solitonic ``constituent quark''
model \ct{const-quark}:

(a) It is charged (the overall charge $Q$ is non-zero \rf{Q-sol}).


(b) It carries non-zero ``color'' flux to infinite -- because of the non-zero
background constant radial electric field \rf{background-el-field}.

\subsection{Matching ``Kinetic Vacuum'' to Standard Constant ``Dilaton''
Vacuum in $(-)$ flat region}

Using the same procedure above we can construct the matching of the 
``kinetic vacuum'' \rf{F-kinetic-minus}-\rf{vp-vac-minus} in $(-)$ flat region 
of the effective scalar potential given by de Sitter metric 
\rf{deSitter-type}-\rf{dS-SdS-def} in the corresponding interior region ($r < r_{(-)}$)
below the de Sitter horizon $r_{(-)}$ \rf{r-minus}, with effective cosmological
constant \rf{CC-eff-minus}, to a static spherically 
symmetric configuration containing the standard constant ``dilaton'' vacuum 
\rf{vp-match-minus} in the outer region ($r > r_{(-)}$) beyond the de Sitter horizon.

In the special case $\xi=0$ ($b=0$), as already noted in \rf{simplify-minus} above,
the expressions \rf{F-kinetic-minus}-\rf{U-kinetic-minus} and \rf{CC-eff-minus} 
in the $(-)$ flat region precisely coincide with the corresponding expressions
\rf{vac-kinetic-plus}-\rf{CC-eff-plus}, \rf{r-plus} in the $(+)$ flat region.
Taking into account \rf{simplify-minus} and the explicit form of $e_{(-)}, f_{(-)}$
\rf{e-f-minus} versus $e_{(+)}, f_{(+)}$ \rf{e-f-plus} and repeating the
same steps as in subsection 5.1 we obtain in the special case $b=0$ a completely 
analogous solution
for the matching of interior ``kinetic vacuum'' \rf{F-kinetic-minus}-\rf{vp-vac-minus} 
corresponding to the $(-)$ flat region of the effective scalar potential
with the exterior region with the same generalized Reissner-Nordstr{\"o}m-de
Sitter geometry carrying a non-zero constant radial background electric field
as in \rf{metric-out-1}-\rf{F-out} with \rf{Q-sol}-\rf{m-sol} upon 
substitution $\bigl(e_{(+)}, f_{(+)}\bigr) \to \bigl(e_{(-)}, f_{(-)}\bigr)$
and $M_{1,2} \to f_{1,2}$. In particular, instead of \rf{coupling-rel} we now obtain 
the following relation among the parameters of the model:
\be
1+\eps \frac{f_1^2}{f_2} - 3 \eps f_0^2 e^2 = 0 \; .
\lab{coupling-rel-minus}
\ee
For the energy densities inequality instead of \rf{rho-out-in} we now have:
\be
\rho_{\rm out} (r) 
\leq \frac{1}{4\eps\chi_2} \Bigl\lb 1 - 
\frac{1}{1+\eps f_1^2/f_2 + \eps e^2 f_0^2}\Bigr\rb 
< \rho_{\rm in} = \frac{1}{4\eps\chi_2} \; .
\lab{rho-out-in-minus}
\ee

Thus, in the special case $\xi=0$ ($b=0$) we obtain a gravitational bag-like 
configuration on the $(-)$ flat region of the effective scalar potential 
with the same properties as those of the gravitational bag solution in 
subsection 5.1 above --  properties (i)-(ii) and (a)-(b).

One can straightforwardly extend the above solution to the general case of
the parameter $b \neq 0$ (\textsl{i.e.},  $\xi \neq 0$). However, the explicit
expressions for the parameters $Q$ and $m$ in the exterior generalized 
Reissner-Nordstr{\"o}m-de Sitter metric carrying a non-zero constant radial 
background electric field as well as the generalization of relation 
\rf{coupling-rel-minus} among the theory's parameters become algebraically
much more complicated. In particular, now the mass parameter $m \neq 0$. 
Moreover, in the general case of $b \neq 0$ ($\xi \neq 0$) the vacuum value of 
the gauge field \rf{F-kinetic-minus}) in the inner finite volume space region 
below the de Sitter horizon ($r < r_{(+)}$) is non-vanishing, thus again
implying confinement dynamics of charge particles as in the outer space region 
beyond the de Sitter horizon ($r > r_{(-)}$ \rf{r-minus}) where we have 
non-zero constant radial background electric field 
$|\vec{E}_{\rm background}| = \frac{1}{\sqrt{2}}e^2_{(-)} f_{(-)}$
(the counterpart of \rf{background-el-field}). On the other hand, according to 
\rf{rho-out-in-minus} the vacuum energy density $\rho_{\rm in}$ 
in the inner finite volume space region is larger than the energy density 
$\rho_{\rm out}$ in the outside region. 

Therefore, in the general case $b \neq 0$ ($\xi \neq 0$) the ``vacuum-like'' solution 
describing matching of ``kinetic vacuum'' to standard constant ``dilaton'' vacuum in 
the $(-)$ flat region of the effective scalar potential is a gravitational bag-like 
solution which shares some of the properties of the ``constituent quark''
model (properties (a)-(c) in subsection 5.1 above), however, it does {\em not at all}
mimic the properties (i)-(iii) of MIT bag unlike the
gravitational bag-like solution of subsection 5.1.

\section{Conclusions}

In the present paper we have constructed a new kind of gravity-matter theory
coupled to a nonstandard nonlinear gauge theory with the following
noteworthy features:

\begin{itemize}
\item
Instead of the canonical Riemannian spacetime volume-form (generally covariant 
integration measure density in terms of $\sqrt{-g}$), to construct the action 
of the model we are employing two different and independent non-Riemannian spacetime
volume-forms defined in terms of two auxiliary antisymmetric tensor gauge
fields of maximal rank.
\item
The action of our model contains apart from the standard Einstein-Hilbert $R$
and Maxwell gauge field $-F^2$ Lagrangian terms also scalar ``dilaton'' parts with a
non-canonical kinetic term, as well as additional $R^2$ and a
``square-root'' Maxwell term $\sqrt{-F^2}$. The specific form of our action
is dictated by the requirement of global Weyl-scale invariance.
\item
Solving the equations of motion for the auxiliary antisymmetric tensor gauge
fields building up the two non-Riemannian spacetime volume-forms introduces
several arbitrary integration constants, some of them spontaneously breaking
the global Weyl-scale symmetry of the initial action.
\item
The physical meaning of the above arbitrary integration constants is
revealed within the canonical Hamiltonian formalism, Namely, these
integration constants turn out to be conserved Dirac-constrained canonical
momenta conjugated to some of the components of the auxiliary antisymmetric tensor gauge
fields of maximal rank, the latter turning out to be essentially pure gauge
non-propagating degrees of freedom.
\item 
After passing to the physical ``Einstein frame'' thanks to the appearance of
the above arbitrary integration constants we obtain a remarkable  effective matter
Lagrangian of quadratic ``k-essence'' type. First, the latter contains an effective scalar
``dilaton'' potential of a very interesting form possessing two infinitely large flat
regions for large negative and large positive ``dilaton'' $\vp$ values. Second, all
the remaining terms in the ``k-essence'' matter Lagrangian appear multiplied
by nontrivial ``dilaton''-dependent coefficient functions, including
nontrivial effective gauge coupling constants running with $\vp$.
\item
We study systematically the static spherically symmetric ``vacuum''-like
configurations corresponding to each of the flat regions of the effective
scalar potential.
\item
First, we find two globally existing in space phases corresponding to the
standard constant ``dilaton'' vacuum values either in either of the two
infinitely large flat regions. These both phases describe confinement since
in both cases the vacuum value of the nonlinear gauge field is non-zero.
\item
Further, we find two localized in space (in ``bubbles'') deconfining (confinement
free) phases corresponding to the so called ``kinetic dilaton vacuum'' (when
the quadratic ``k-essence'' effective action in the Einstein frame is extremized 
w.r.t. $X$ - the ``dilaton'' kinetic term), where the vacuum value of the nonlinear 
gauge field vanishes. In one of these deconfining phases the ``dilaton''
lies on the $(+)$ flat region of the effective scalar potential and in the
second one it belongs to the $(-)$ flat region with the additional
restriction on the Lagrangian parameter $b=0$. On the other hand, in the
generic case of $b\neq 0$ the ``kinetic dilaton vacuum'' on the $(-)$ flat
region describes a localized confining phase since the vacuum value of the nonlinear
gauge field is again non-zero there.
\item
The localized deconfining phases inside the ``bubbles'' coexist with outside
configurations corresponding to standard constant ``dilaton'' vacuums and
nontrivial nonlinear gauge fields carrying non-zero constant radial
background electric field 
The energy density inside the ``bubbles'' is larger that the
outside energy density. Thus, the full solution inside and outside the
``bubbles'' is a gravitational bag-like solution mimicking some of the basic
properties of the MIT bag \ct{MIT-bag-1,MIT-bag-2}.
\item
On the other hand, the analogy of the above gravitational bag with the MIT
bag is only partial one. The present gravitational bag possess other
properties (overall charge, 
carrying non-zero ``color'' flux to infinity) which resemble some of the main 
properties of the solitonic ``constituent quark'' model \ct{const-quark}.
\end{itemize}








\appendix

\section{Canonical Hamiltonian Treatment of Gravity-Matter Theories with
Non-Riemannian Volume-Forms}

Here we will briefly discuss the application of the canonical Hamiltonian formalism 
to the new gravity-matter model based on two non-Riemannian spacetime volume-forms 
\rf{TMMT+GG}. In order to elucidate the proper physical meaning of the arbitrary
integration constants $\chi_2,\, M_1,\, M_2$ \rf{integr-const} encountered
within the Lagrangian formalism's treatment of \rf{TMMT+GG} it is sufficient to
concentrate only on the canonical Hamiltonian structure related to the
auxiliary maximal rank antisymmetric tensor gauge fields 
$A_{\m\n\l}, B_{\m\n\l}, H_{\m\n\l}$ and their respective conjugate momenta.

For convenience let us introduce the following short-hand notations for the
field-strengths \rf{Phi-1-2}, \rf{Phi-H} of the auxiliary 3-index antisymmetric gauge 
fields (the dot indicating time-derivative): 
\br
\P_1 (A) = \Adot + \pa_i A^i \quad, \quad 
A = \frac{1}{3!} \vareps^{ijk} A_{ijk} \;\; ,\;\;
A^i = - \h \vareps^{ijk} A_{0jk} \; ,
\lab{A-can} \\
\P_2 (B) = \Bdot + \pa_i B^i \quad, \quad 
B = \frac{1}{3!} \vareps^{ijk} B_{ijk} \;\; ,\;\;
B^i = - \h \vareps^{ijk} B_{0jk} \; ,
\lab{B-can} \\
\P (H) = \Hdot + \pa_i H^i \quad, \quad 
H = \frac{1}{3!} \vareps^{ijk} H_{ijk} \;\; ,\;\;
H^i = - \h \vareps^{ijk} H_{0jk} \; ,
\lab{H-can}
\er
Also we will use the short-hand notation:
\be
{\wti L}^{(1)} (u,\udot) \equiv R + L^{(1)} \quad ,\quad
{\wti L}^{(2)} (u,\udot) \equiv L^{(2)} + \eps R^2 \; ,
\lab{L-tilde}
\ee
where $L^{(1,2)}$ are as in \rf{L-1}-\rf{L-2} and where $(u,\udot)$ collectively denote 
the set of the basic gravity-matter canonical variables 
$(u)=\bigl(g_{\m\n}, \vp, A_\m \bigr)$ and their respective velocities.

For the pertinent canonical momenta conjugated to \rf{A-can}-\rf{H-can} we have:
\br
\pi_A = {\wti L}_1 (u,\udot) \;\; ,\;\;
\pi_B = {\wti L}^{(2)} (u,\udot) + \frac{1}{\sqrt{-g}}(\Hdot + \pa_i H^i) \; ,
\nonu \\
\pi_H = \frac{1}{\sqrt{-g}}(\Bdot + \pa_i B^i) \; ,
\lab{can-momenta-aux}
\er
and:
\be
\pi_{A^i} = 0 \quad,\quad \pi_{B^i} = 0 \quad,\quad \pi_{H^i} = 0 \; .
\lab{can-momenta-zero}
\ee
The latter imply that $A^i, B^i, H^i$ will in fact appear as Lagrange multipliers
for certain first-class Hamiltonian constraints 
(see Eqs.\rf{pi-A-const}-\rf{pi-B-pi-H-const} below). 
For the canonical momenta conjugated to the basic gravity-matter canonical variables 
we have (using last relation \rf{can-momenta-aux}):
\be
p_u = (\Adot + \pa_i A^i) \frac{\pa}{\pa \udot} {\wti L}_1 (u,\udot) + 
\pi_H \sqrt{-g} \frac{\pa}{\pa \udot} L^{(2)} (u,\udot) \; .
\lab{can-momenta-u}
\ee

Now, relations \rf{can-momenta-aux} and \rf{can-momenta-u} allow us to
obtain the velocities $\udot,\,\Adot,\,\Bdot,\,\Hdot$ as functions
of the canonically conjugate momenta $\udot = \udot (u,p_u,\pi_A,\pi_B,\pi_H)$
\textsl{etc.} (modulo some Dirac constraints among the basic gravity-matter
variables due to general coordinate and gauge invariances). Taking into account
\rf{can-momenta-aux}-\rf{can-momenta-zero} (and the short-hand notations
\rf{A-can}-\rf{L-tilde}) the canonical Hamiltonian corresponding to \rf{TMMT+GG}:
\br
\cH = p_u \udot + \pi_A \Adot + \pi_B \Bdot + \pi_H \Hdot -
(\Adot + \pa_i A^i) {\wti L}_1 (u,\udot) 
\nonu \\
- \pi_H \sqrt{-g} \Bigl\lb {\wti L}^{(2)}(u,\udot) + 
\frac{1}{\sqrt{-g}}(\Hdot + \pa_i H^i) \Bigr\rb
\lab{can-hamiltonian}
\er
acquires the following form as function of the canonically conjugated variables
(here $\udot = \udot (u,p_u,\pi_A,\pi_B,\pi_H)$):
\br
\cH = p_u \udot - \pi_H \sqrt{-g} {\wti L}^{(2)}(u,\udot)
\nonu \\
+ \sqrt{-g} \pi_H \pi_B - \pa_i A^i \pi_A - \pa_i B^i \pi_B - \pa_i H^i \pi_H \; .
\lab{can-hamiltonian-final}
\er
From \rf{can-hamiltonian-final} we deduce that indeed $A^i, B^i, H^i$ are Lagrange 
multipliers for the first-class Hamiltonian constraints:
\be
\pa_i \pi_A = 0 \;\; \to\;\; \pi_A = - M_1 = {\rm const} \; ,
\lab{pi-A-const}
\ee
and similarly:
\be
\pi_B = - M_2 = {\rm const} \quad ,\quad \pi_H = \chi_2 = {\rm const} \; ,
\lab{pi-B-pi-H-const}
\ee
which are the canonical Hamiltonian counterparts of Lagrangian constraint
equations of motion \rf{integr-const}.

Thus, the canonical Hamiltonian treatment of \rf{TMMT+GG} reveals the meaning
of the auxiliary 3-index antisymmetric tensor gauge fields
$A_{\m\n\l},\, B_{\m\n\l},\, H_{\m\n\l}$ -- building blocks of
the non-Riemannian spacetime volume-form formulation of the modified gravity-matter
model \rf{TMMT+GG}. Namely, the canonical momenta $\pi_A,\, \pi_B,\, \pi_H$ 
conjugated to the ``magnetic'' parts $A,B,H$ \rf{A-can}-\rf{H-can}
of the auxiliary 3-index antisymmetric tensor gauge fields are constrained
through Dirac first-class constraints \rf{pi-A-const}-\rf{pi-B-pi-H-const}
to be constants identified with the arbitrary 
integration constants $\chi_2,\, M_1,\, M_2$ \rf{integr-const} arising within the 
Lagrangian formulation of the model. The canonical momenta 
$\pi_A^i,\, \pi_B^i,\, \pi_H^i$ conjugated to the ``electric'' parts $A^i,B^i,H^i$ 
\rf{A-can}-\rf{H-can} of the auxiliary 3-index antisymmetric tensor gauge field
are vanishing \rf{can-momenta-zero} which makes the latter canonical Lagrange 
multipliers for the above Dirac first-class constraints.

\section{``Cornell''-Type Confining Potential in Curved Spacetime}
Here we will follow the steps of the derivation in Ref.~\refcite{GG-2} of
effective ``Cornell''-type confining potential \ct{cornell-1,cornell-2,cornell-3} between 
quantized charged fermions based on the general formalism \ct{bunster-77} for quantization 
within the canonical Hamiltonian approach a'la Dirac of truncated gauge and gravity theories 
by imposing explicitly spherical symmetry on the pertinent Lagrangian action. 

In the present case the corresponding nonlinear gauge field action:
\be
S = \int d^4 x\,\sqrt{-g} \Bigl\lb L(F^2) + A_\m J^\m \Bigr\rb \quad, \quad
L(F^2) = - \frac{1}{4} F^2 - \frac{f_0}{2}\sqrt{-F^2} 
\lab{NL-action}
\ee
yields equations of motion:
\be
\pa_\n \Bigl(\sqrt{-g}4 L^\pr (F^2) F^{\m\n}\Bigr) + \sqrt{-g} J^\m = 0
\quad ,\quad L^\pr (F^2) = -\frac{1}{4}\Bigl( 1 - \frac{f_0}{\sqrt{-F^2}}\Bigr) \; ,
\lab{NL-eqs}
\ee
whose $\m=0$ component -- the nonlinear ``Gauss law'' constraint equation reads:
\be
\frac{1}{\sqrt{-g}} \pa_i \bigl(\sqrt{-g} D^i\bigr) = J^0 \quad ,\quad 
D^i = \Bigl( 1 - \frac{f_0}{\sqrt{-F^2}}\Bigr) F^{0i} \; ,
\lab{NL-gauss-law}
\ee
with $\vec{D}\equiv (D^i)$ denoting the electric displacement field
nonlinearly related to the electric field $\vec{E}\equiv (F^{0i})$ as in the
last relation \rf{NL-gauss-law}. 

In the special case of nonlinear gauge field theory
\rf{NL-action} there exists a nontrivial vacuum solution $\sqrt{-F^2_{\rm
vac}} = f_0$, which implies simultaneous vanishing of the electric displacement field,
$\vec{D}=0$ meaning zero observed charge, and at the same time nontrivial
electric field. In particular, for static spherically symmetric fields in
static spherically symmetric metric (of the form \rf{deSitter-type} with
general $\cA (r)$) the only surviving component of $F_{\m\n}$ is the nonvanishing radial
component of the electric field $E^r = - F_{0r}$, so that 
$\sqrt{-F^2_{\rm vac}} = \sqrt{2} |\vec{E}|= f_0$. This can be viewed as the simplest
classical manifestation of charge confinement: $\vec{D}=0$ and nontrivial $\vec{E}$.

Here we will employ the canonical Hamiltonian treatment in Ref.~\refcite{bunster-77} 
and will truncate the nonlinear gauge field action to purely spherically
symmetric fields, \textsl{i.e.}, we will take $F_{0r} = \pa_0 A_r - \pa_r A_0$ independent
of the space angles and the rest of the components of $F_{\m\n}$ being zero. 
The action of the truncated theory reads:
\be
S_{\rm truncated} = \int dt \int dr 4\pi r^2 \Bigl\lb \h F_{0r}^2 - 
\frac{f_0}{\sqrt{2}}|F_{0r}| + A_0 J^0 + A_r J^r \Bigr\rb \; .
\lab{NL-action-truncated}
\ee
Note that in \rf{NL-action-truncated} there is no explicit dependence on the
Riemannian metric coefficient $\cA (r) = - g_{00} = 1/g_{rr}$. 
It is now straightforward to apply the canonical Hamiltonian quantization
procedure to \rf{NL-action-truncated} within the Dirac formalism for constrained dynamical
systems (\textsl{e.g.} Ref.~\refcite{henneaux-bunster}). Obviously, in the
case of de Sitter spacetime the radial coordinate $r$ must be restricted to
vary up to the de Sitter horizon radius $r_H$.

The canonically conjugated momenta w.r.t. $A_0$ and $A_r$ read:
\be
\Pi^0 = 0 \quad ,\quad \Pi^r = 4\pi r^2 \Bigl( F_{0r} - \frac{f_0}{\sqrt{2}}\Bigr) \; ,
\lab{canon-momenta}
\ee
where the first one $\Pi^0 = 0$ is the standard primary Dirac constraint known in any
gauge theory of Yang-Mills type. For the density of the canonical Hamiltonian one obtains:
\be
\cH = \frac{1}{8\pi r^2} \(\Pi^r\)^2 + \frac{f_0}{\sqrt{2}}\Pi^r +
\pi r^2 f_0^2 - A_r J^r + \Pi^r \pa_r A_0 - J^0  A_0 \; .
\lab{canon-H-0}
\ee
Henceforth, for simplicity we will consider the case with no matter current $J^r = 0$.
Time-preservation of the primary constraint $\Pi^0 = 0$, \textsl{i.e.},
$\frac{d}{dt}\Pi^0 = \Bigl\{ \Pi^0 ,\cH \Bigr\}_{\rm PB} = 0$ yields the
standard secondary Dirac constraint -- the ``Gauss law'' constraint:
\be
\Phi_1 (r) \equiv \pa_r \Pi^r + J^0 = 0 \; .
\lab{gauss-law}
\ee

Thus, one has to Dirac-canonically quantize the theory with canonical
Hamiltonian:
\be
H = \int dr \Bigl\lb \frac{1}{8\pi r^2} \bigl(\Pi^r\bigr)^2 + \frac{f_0}{\sqrt{2}}\Pi^r +
\pi r^2 f_0^2 \Bigr\rb
\lab{canon-H-1}
\ee
and with two first-class a'la Dirac constraints $\Phi_{0,1}=0$ ($\Phi_0 \equiv \Pi^0 = 0$ 
and $\Phi_1 = 0$ as in \rf{gauss-law}), which have to be supplemented by two canonically 
conjugate gauge-fixing conditions $\chi_{0,1}$. Since $A_0$ and its
conjugate momentum $\Pi^0 = 0$ do not mix with the rest of the canonical
variables they have no impact on the pertinent {\em Dirac brackets} between
$A_r$ and $\Pi^r$ to be promoted to quantum operator commutators upon quantization.
Thus we only need to choose an appropriate gauge fixing condition for the ``Gauss law'' 
constraint \rf{gauss-law}, which we can take in the form:
\be
\chi_1 (r) \equiv \int_{C(r)} dz^\l A_\l (z) \; . 
\lab{Psi-1}
\ee
Here $\int_{C(r)}$ is path integral along a spacelike geodesic $x^\l = x^\l (\xi)$ ending at 
the spacetime point with radial coordinate $r$. In particular, for the interior de Sitter
region ($r \leq r_H$) this spacelike geodesic can be taken in the form:
\be
t(\xi) = t = {\rm const} \quad ,\quad r(\xi) = r_H \sin (\xi/r_H) \;\; , \;\;
0 \leq \xi \leq \xi_{\rm fin} \leq r_H \frac{\pi}{2} \;\; , \;\; r(\xi_{\rm fin}) = r \; ,
\lab{dS-curve}
\ee
where $\xi$ is the de Sitter proper distance parameter, so that:
\be
\chi_1 (r) \equiv \int_0^r dz A_r (z) \quad,\quad
\Bigl\{\Phi_1 (r),\chi_1 (r^\pr)\Bigr\}_{\rm PB} = \d (r-r^\pr) \; .
\lab{PB-Phi-Psi}
\ee
Note that here and below $\d (r-r^\pr)$ denotes the Dirac delta-function on the half-line
(both $r, r^\pr > 0$).

It is now straightforward to calculate the Dirac bracket between the
canonically conjugate pair given by:
\br
\Bigl\{ A_r (r), \Pi^r (r^\pr)\Bigr\}_{\rm DB} = 
\Bigl\{ A_r (r), \Pi^r (r^\pr)\Bigr\}_{\rm PB} \phantom{aaaa}
\nonu \\
- \int\int dr^{\pr\pr} dr^{\pr\pr\pr}
\Bigl\{ A_r (r), \Phi_1 (r^{\pr\pr})\Bigr\}_{\rm PB}
\Bigl\{\Phi_1 (r^{\pr\pr}),\chi_1 (r^{\pr\pr\pr})\Bigr\}^{-1}_{\rm PB}
\Bigl\{\chi_1 (r^{\pr\pr\pr}),\Pi^r (r^\pr)\Bigr\}_{\rm PB} \; , \phantom{aaaa}
\lab{DB-def}
\er
by using the standard Poisson bracket 
$\Bigl\{ A_r (r), \Pi^r (r^\pr)\Bigr\}_{\rm PB} = \d (r-r^\pr)$, which yields:
\be
\Bigl\{ A_r (r), \Pi^r (r^\pr)\Bigr\}_{\rm DB} = 2 \d (r-r^\pr) \; .
\lab{DB}
\ee
Upon canonical quantization \rf{DB} becomes:
\be
\llb {\widehat\Pi}^r (r), {\widehat A}_r (r^\pr) \rrb = 2i \d (r-r^\pr)
\quad, \;\; {\mathrm i.e.}\;\; {\widehat\Pi}^r (r) = -2i\d/\d A_r (r) \; .
\lab{Dirac-CCR}
\ee

Now, following Ref.~\refcite{GG-2} we consider a gauge invariant quantum state of two
oppositely charged ($\pm e_0$) fermions located at $r=0$ and $r=L$, respectively,
explicitly given by:
\be
\left | \Phi \rrangle \equiv \left | {\bar\Psi}(L) \Psi (0) \rrangle =
{\bar\Psi}(L) \exp\Bigl\{ie_0 \int_0^L dz A_r (z)\Bigr\}\Psi (0) \left | 0 \rrangle
\; .
\lab{two-fermion-state}
\ee
The average of the quantized canonical Hamiltonian \rf{canon-H-1} in this state 
\rf{two-fermion-state}, where now $\Pi^r (r)$ will act on $A_r (r)$ according to 
\rf{Dirac-CCR}:
\be
\langle \Phi \vert {\widehat H} \vert \Phi\rangle 
\equiv V_{\rm eff} (L)
\lab{V-eff}
\ee
can be viewed as effective potential between the quantized fermionic pair generated by 
the nonlinear gauge field theory containing the ``square-root'' Maxwell term
\rf{NL-action}.

Using \rf{Dirac-CCR} one calculates:
\br
\Bigl\lb {\widehat\Pi}^r (r) , ie_0 \int_0^L dz A_r (z) \Bigr\rb = 
2 e_0 \th (L-r) \; , 
\lab{commute-1} \\
\Bigl\lb \Bigl\lb \({\widehat\Pi}^r (r)\)^2 , ie_0 \int_0^L dz A_r (z) \Bigr\rb ,
ie_0 \int_0^L dz A_r (z) \Bigr\rb = 8 e^2_0 \th (L-r) \; , 
\lab{commute-2}
\er
where $\th (r-r^\pr)$ denotes the step-function on the half-line (both $r, r^\pr > 0$).
Plugging \rf{commute-1}-\rf{commute-2} into \rf{V-eff} we obtain:
\be
V_{\rm eff} (L) = - \frac{e_0^2}{2\pi} \frac{1}{L} + e_0 f_0 \sqrt{2}\, L +
\bigl( L{\rm -independent} ~{\rm const} \bigr) \; ,
\lab{cornell-type}
\ee
which has precisely the form of the ``Cornell'' potential
\ct{cornell-1,cornell-2,cornell-3}.

\section*{Acknowledgments.} 
We gratefully acknowledge support of our collaboration through the academic exchange 
agreement between the Ben-Gurion University and the Bulgarian Academy of Sciences.
E.N. and S.P. are partially supported by Bulgarian NSF grant {\em DFNI T02/6}.
S.P. has received partial support from European COST action MP-1210.
E.N. has received partial support from European COST action MP-1405.


\end{document}